\documentclass[reprint,superscriptaddress,longbibliography, prl]{revtex4-1}
\usepackage{color}
\usepackage{graphicx}
\usepackage{hyperref}
\usepackage{amsmath}
\usepackage{amssymb}
\usepackage{array}
\usepackage{dcolumn}
\usepackage{color}
\usepackage{bm}
\usepackage{multirow}
\usepackage{makecell}
\usepackage{subfigure}
\usepackage{soul}
\usepackage{color}
\usepackage{tabularx}
\newcolumntype{C}{>{\centering\arraybackslash}p{3.8cm}}

\hypersetup{
    pdfnewwindow=true,    
    colorlinks=true,      
    linkcolor=blue,       
    citecolor=blue,       
    filecolor=blue,       
    urlcolor=blue         
}

\definecolor{forestgreen}{rgb}{0.13, 0.55, 0.13}

\begin{document}

\title{Constraints on Axion-like Particles from a Hard $X$-ray Observation of Betelgeuse}

\author{Mengjiao Xiao}
 \email{mjxiao@mit.edu}
\author{Kerstin M. Perez}
  \email{kmperez@mit.edu}
\affiliation{Department of Physics, Massachusetts Institute of Technology, Cambridge, Massachusetts 02139, USA}
\author{Maurizio Giannotti}
  \email{mgiannotti@barry.edu}
  \affiliation{Physical Sciences, Barry University, 11300 NE 2nd Ave., Miami Shores, Florida 33161, USA}
\author{Oscar Straniero}
  \email{oscar.straniero@inaf.it}
  \affiliation{INAF, Osservatorio Astronomico d’Abruzzo, 64100 Teramo, Italy}
\author{Alessandro Mirizzi}
  \email{alessandro.mirizzi@ba.infn.it}
  \affiliation{Dipartimento Interateneo di Fisica “Michelangelo Merlin”, Via Amendola 173, 70126 Bari, Italy}
  \affiliation{Istituto Nazionale di Fisica Nucleare - Sezione di Bari, Via Orabona 4, 70126 Bari, Italy}
\author{Brian W. Grefenstette}
  \email{bwgref@srl.caltech.edu}
  \affiliation{Cahill Center for Astrophysics, 1216 E. California Blvd, California Institute of Technology, Pasadena, California 91125, USA}
\author{Brandon M. Roach}
  \email{roachb@mit.edu}
  \affiliation{Department of Physics, Massachusetts Institute of Technology, Cambridge, Massachusetts 02139, USA}
\author{Melania Nynka}
  \email{mnynka@mit.edu}
  \affiliation{MIT Kavli Institute for Astrophysics and Space Research, 77 Massachusetts Avenue, Cambridge, MA 02139, USA}
\date{\today}

\begin{abstract}
We use the first observation of Betelgeuse in hard $X$-rays to perform a novel search for axion-like particles (ALPs). Betelgeuse is not expected to be a standard source of $X$-rays, but light ALPs produced in the stellar core could be converted back into photons in the Galactic magnetic field, producing a detectable flux that peaks in the hard $X$-ray band ($E_\gamma>10\mathrm{\,keV}$). Using a 50\,ks observation of Betelgeuse by the \emph{NuSTAR} satellite telescope, we find no significant excess of events above the expected background. Using models of the regular Galactic magnetic field in the direction of Betelgeuse, we set a 95\% C.L. upper limit on the ALP-photon coupling of ${g_{a\gamma}<(0.5-1.8)\times10^{-11}}$ \textrm{GeV}$^{-1}$ (depending on magnetic field model) for ALP masses ${m_{a}<(5.5-3.5) \times10^{-11}}$ eV.
\end{abstract}

\maketitle

\nocite{*}

\noindent \emph{Introduction.}---Axion-like particles (ALPs) are ultralight pseudoscalar bosons with a two-photon vertex $g_{a\gamma}$, predicted by several extensions of the Standard Model (see~\cite{Jaeckel:2010ni, DiLuzio:2020wdo} for a recent review). In the presence of an external magnetic field, the {$g_{a\gamma}$} coupling leads to the phenomenon of photon-ALP mixing~\cite{Raffelt:1987im}. This effect is exploited by several ongoing and upcoming ALP search experiments (see~\cite{Irastorza:2018dyq,DiLuzio:2020wdo,Sikivie:2020zpn} for recent reviews).

The photon-ALP coupling would also cause ALPs to be produced in stellar plasmas via the Primakoff process \cite{Raffelt:1985nk}. Therefore astrophysical observations offer unique sensitivity to ALPs. Analyses of the lifetime of helium-burning stars in globular clusters have excluded the couplings $g_{a\gamma}>6\times10^{-11}\mathrm{\,GeV}^{-1}$ (95\% confidence level, C.L.) for $m_a\lesssim30\,\ \mathrm{\,keV}$~\cite{Ayala:2014pea,Straniero:2015nvc,Carenza:2020zil}. This is the strongest bound on the ALP-photon coupling in a wide mass range. A comparable bound, $g_{a\gamma}>6.6\times10^{-11}\mathrm{\,GeV}^{-1}$ (95\% C.L.) for $m_{a}\lesssim0.02 \,\  \mathrm{\,eV}$, was derived by the CAST experiment, which searches for ALPs produced in the Solar core that are re-converted into $X$-rays in a large laboratory magnetic field \cite{cast:2017ftl}.

\begin{figure}[htb!]
\centering
\includegraphics[width=\columnwidth]{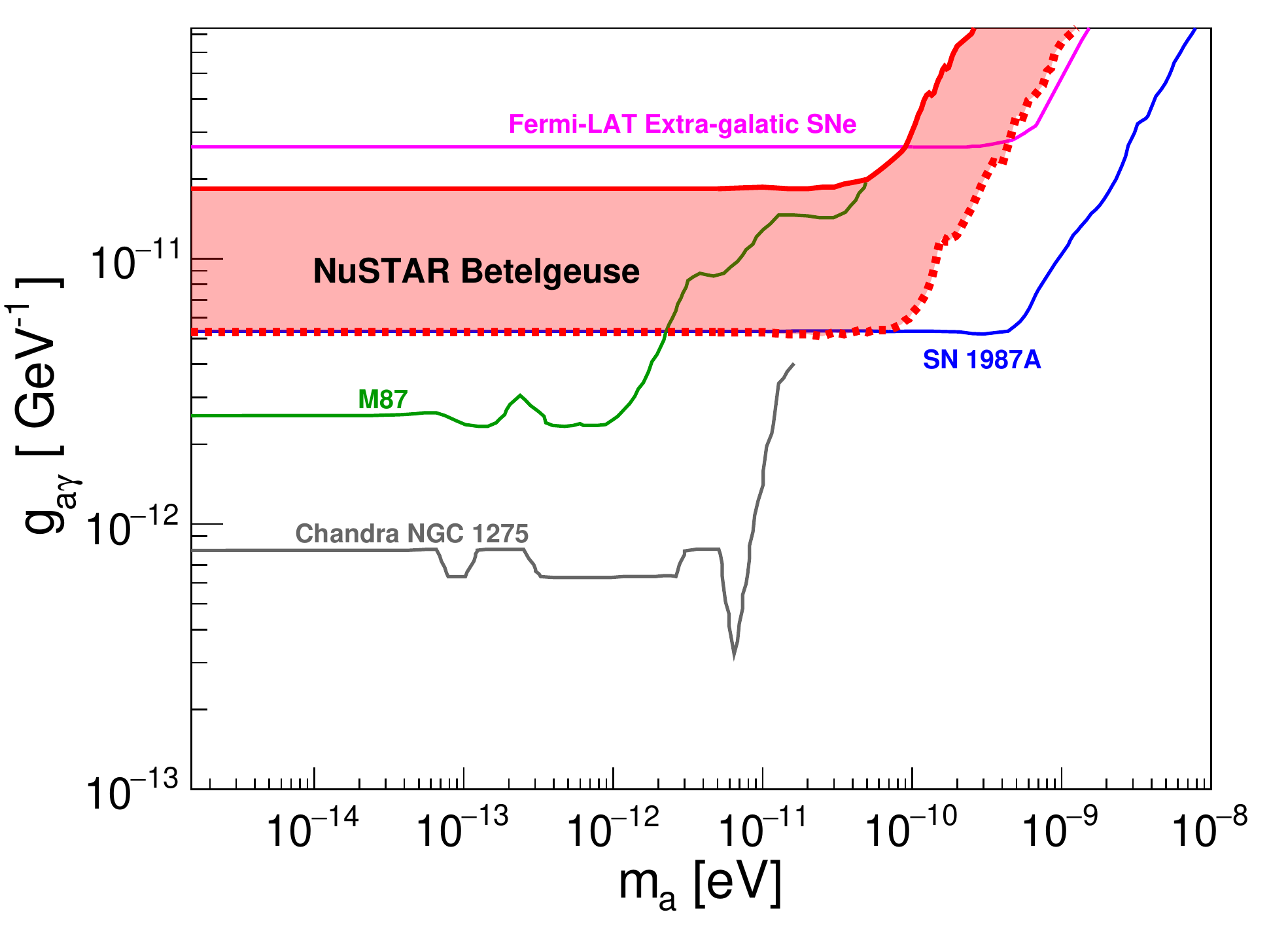}
\caption{Astrophysical constraints on the ALP-photon coupling in relation to the Betelgeuse bound from the present work: SN 1987A \cite{payez:2014xsa}, Fermi-LAT extra-galactic SNe \cite{Meyer:2020vzy}, M87 \cite{m87:2017yvc}, and Chandra NGC 1275 \cite{chandra:2019uqt}.}
\label{fig:upperlimts_lowmass}
\end{figure}
At lower ALP masses, as shown in Fig.~\ref{fig:upperlimts_lowmass}, more stringent bounds are derived from supernovae and galaxy cluster observations. A recent reanalysis of the SN 1987A limit, derived from the non-observation of gamma-rays induced by conversion in the Galactic magnetic field of ALPs produced in the supernova core, excludes ${g_{a\gamma}>5.3\times10^{-12}\mathrm{\,GeV}^{-1}}$ for $m_{a}<4.4\times10^{-10}\mathrm{\,eV}$ (95\% C.L.)~\cite{payez:2014xsa}. A search for gamma-ray bursts from extra-galactic supernovae with the Fermi Large Area Telescope (LAT) has yielded a weaker limit, $g_{a\gamma} \lesssim 2.6 \times 10^{-11}\mathrm{\,GeV}^{-1}$ for ALP masses $m_{a} \lesssim 3 \times 10^{-10}\mathrm{\,eV}$ \cite{Meyer:2020vzy}, under the assumption of at least one supernova occurring in the LAT field of view. However, the SN 1987A bound has been questioned due to uncertainties on the response of the GRS instrument on the SMM satellite~\cite{Forrest:1980}, and both of these bounds are subject to uncertainties due to the modeling of core-collapse supernovae and Galactic magnetic fields. The most stringent limits to date on low-mass ALPs come from the search for $X$-ray spectral distortions in the active galactic nucleus NGC 1275 at the center of the Perseus cluster, induced by conversion in the intra-cluster medium.  
These exclude ${g_{a\gamma}>(6-8)\times10^{-13}\mathrm{\,GeV}^{-1}}$ (99.7\% C.L.) for $m_{a}<10^{-12}\mathrm{\,eV}$ \cite{chandra:2019uqt}. However, these results could be weakened by several orders of magnitude if the intra-cluster magnetic field has been mis-modeled~\cite{Libanov:2019fzq}.

Here, we exploit the fact that Betelgeuse, a nearby red supergiant star, provides an excellent laboratory for ALPs, as proposed in a seminal paper by E.~Carlson \cite{Carlson:1995wa}. Betelgeuse ($\alpha$ Orionis, spectral type M2Iab) is an ideal candidate for ALP searches, as it $(i)$ has a hot core, and thus is potentially a copious producer of ALPs that, after re-conversion in the Galactic magnetic field, produces a photon signal peaked in the hard $X$-ray range, and $(ii)$ is in region of Hertzsprung-Russel diagram where no stable corona is expected, and thus has essentially zero standard astrophysical $X$-ray background~\cite{PossonBrown:2006zr}.
This basic idea can be extended to other stellar objects, such as clusters of hot, young stars \cite{ben2020}.
Betelgeuse has the additional advantage that it is nearby, at a distance $d{\sim}200\mathrm{\,pc}$~\cite{Harper_2008,Harper_2017}, and thus in a region of the local magnetic field that is relatively easier to constrain with future  observations.

In this work, we use a dedicated observation of Betelgeuse by the \emph{NuSTAR} satellite telescope to derive limits on the ALP-photon coupling $g_{a\gamma}<(0.5-1.8)\times10^{-11}\mathrm{\,GeV}^{-1}$ (95\% C.L.) for masses $m_{a}<(5.5-3.5) \times10^{-11}\mathrm{\,eV}$ (depending on the assumed value of the regular Galactic magnetic field). We derive a range of ALP spectra expected from Betelgeuse, depending on its precise evolutionary stage. We then derive the expected $X$-ray spectra after conversion in the regular Galactic magnetic field, and compare to our spectral measurement in the energy range 3--79 keV. For low-mass ALPs, these limits are a factor of $\sim$3 deeper than the limits from CAST, and are comparable to the limits expected from next-generation ALP experiments such as ALPS-II~\cite{alps2:2013ywa} and BabyIAXO~\cite{Armengaud:2019uso, DiVecchia:2019ejf}. They are competitive with the constraints from SN 1987A, though with independent sources of systematic uncertainty. We emphasize that the preponderance of astrophysical exclusions that overlap in this region of $g_{a\gamma}-m_{a}$ parameter space, each with a separate modeling assumptions and sources of uncertainty, builds confidence in the robustness of these constraints.

\noindent \textit{ALP and photon fluxes from Betelgeuse.}---In the minimal scenarios, ALPs have only a two-photon coupling, described by the Lagrangian~\cite{Raffelt:1987im}:
\begin{equation}
\label{mr}
\mathcal{L}_{a\gamma}=-\frac{1}{4} \,
F_{\mu\nu}\tilde{F}^{\mu\nu}a= \, {\bf E}\cdot{\bf B}\,a~.
\end{equation}
This interaction allows for ALP production in a stellar medium primarily through the Primakoff process, in which thermal photons are converted into ALPs in the electrostatic field of ions, electrons and protons. The ALP production rate via the Primakoff process in a stellar core can be calculated as (see, e.g.,~\cite{Carlson:1995wa}),
\begin{eqnarray}
\frac{d \dot n_a}{dE}&=&
\frac{g_{a\gamma}^{2}\xi^2\, T^3\,E^2}{8\pi^3\, \left( e^{E/T}-1\right) } \nonumber \\
& &\left[ \left( 1+\frac{\xi^2 T^2}{E^2}\right)  \ln\left(1+\frac{E^2}{\xi^2T^2}\right) -1 \right] \,,
\label{eq:axprod}
\end{eqnarray}
where $E$ is the photon energy, $T$ the temperature, and ${\xi^2={\kappa^2}/{4T^2}}$ with $\kappa$ the inverse of the screening length, introduced by the finite range of electric field surrounding charged particles in the plasma.
Once produced, ALPs can easily escape the star since their mean free path in stellar matter is sufficiently large for the values of mass and coupling we are interested in. The total ALP spectrum can then be obtained by integrating Eq.~(\ref{eq:axprod}) over the volume of the star, $d\dot N_{a}/dE = \int (d\dot n_{a}/dE)dV$, which can be well parametrized by fitting the stellar model (see Supplementary Material for details).

The ALP spectrum thus depends on the physical structure and chemical evolution of the star. Betelgeuse, due to its relative proximity, has fairly well-constrained observed values of luminosity, effective temperature, and metallicity (see Supplementary Material). However, the time until core-collapse is not well known, and as the temperature and density of the core increase as the star approaches supernova, this introduces significant uncertainty on the predicted ALP spectra. We use the Full Network Stellar evolution code (FuNS~\cite{Straniero:2019}) to derive 13 models of the Betelgeuse ALP source spectrum, parametrized by the time until core-collapse, $t_\mathrm{cc}$. These range from an optimistic ALP production scenario ($t_\mathrm{cc}$ = 1.4 yr) to a conservative ALP flux scenario ($t_\mathrm{cc}$ = $1.55\times10^{5}$ yr), as shown in Tab.~\ref{tab:axion_models}.

The interaction in Eq.~(\ref{mr}) may also trigger ALP--photon oscillations in external magnetic fields, such as those found in our galaxy, thus producing a detectable $X$-ray flux. The calculation of the ALP-photon re-conversion probability simplifies if we restrict ourselves to the case in which ${\bf B}$ is homogeneous. The magnetic field of the Galaxy is known to change on scales of $\sim$1\,kpc, corresponding to the arm and inter-arm regions, and between the Galactic disk and halo~\cite{Han2017, Jansson2012, XuHan2019}. Motivated by the relative proximity of Betelgeuse in the Galactic disk, we assume conversion in a homogenous regular magnetic field. Although the magnetic structure of the Galaxy is certainly more complex -- including turbulent fields with coherent lengths $\mathcal{O}$(200\,pc) or smaller~\cite{XuHan2019,Pelgrims2020,Beck2016}, such as due to supernovae, molecular clouds, and our own Local Bubble -- this assumption follows the convention of previous astrophysical bounds, e.g. SN 1987A~\cite{payez:2014xsa}. This allows for consistent comparison between constraints. A detailed treatment of the effect of magnetic field correlations lengths and amplitudes on the conversion probability is deferred to a later dedicated theoretical study.

\begin{figure*}[t]
\centering
\includegraphics[width=2.\columnwidth]{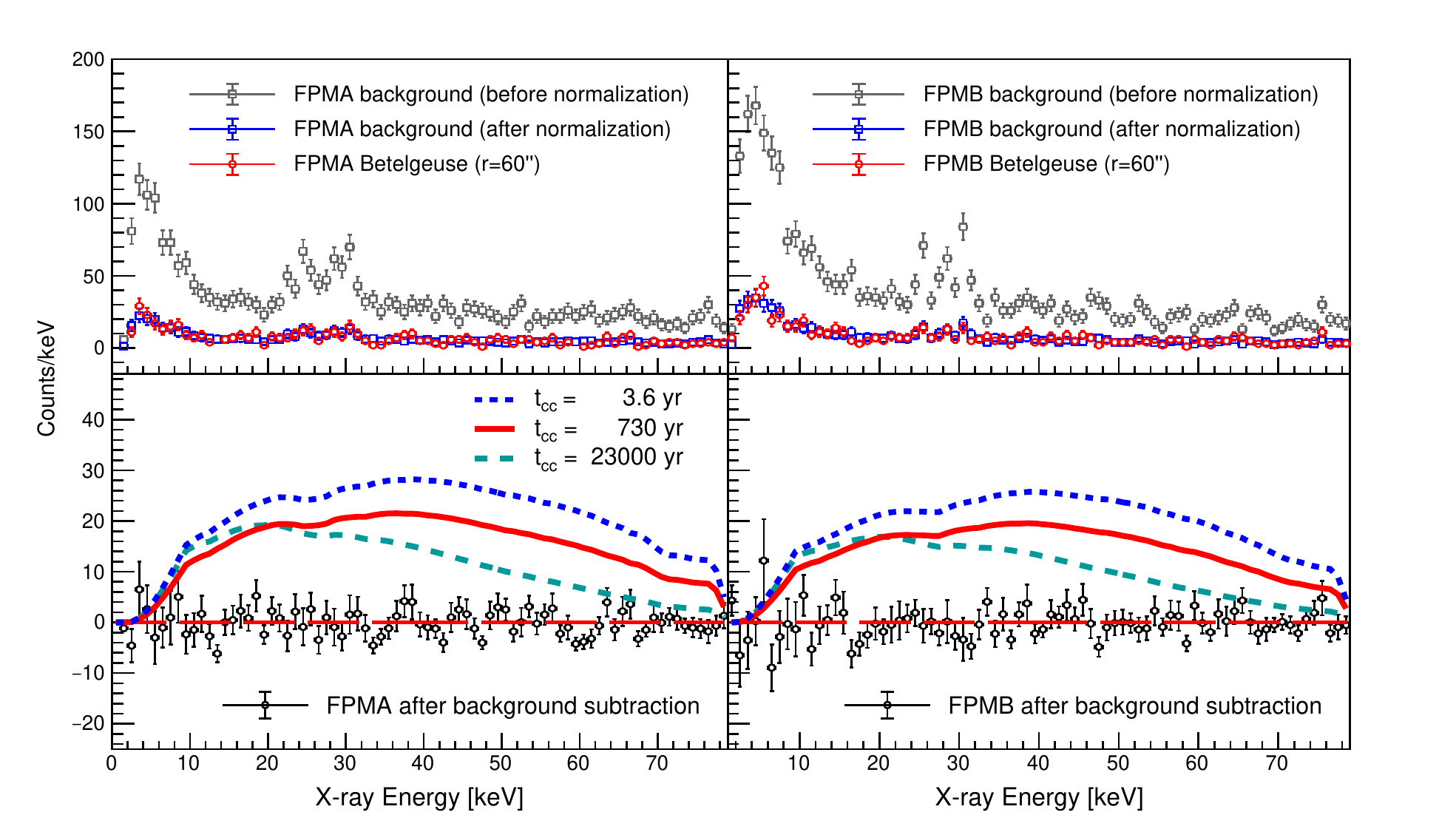}
\caption{\emph{Top:} $X$-ray spectra from FPMA (\emph{left}) and FPMB (\emph{right}) for the Betelgeuse source (red) and background (gray and blue for before and after normalization) regions. The error bars overlaid are the statistic uncertainties ($\sqrt{N}$). \emph{Bottom:} Source spectra after subtracting the normalized background. The error bars are calculated by Sumw2 with ROOT software \cite{Brun:1997pa}. The predicted ALP-produced $X$-ray spectra assuming transverse magnetic field $B_{\rm T}=1.4~\,\mu\rm G$, mass $m_a=10^{-11}\mathrm{\,eV}$ and coupling $g_{a\gamma}=1.5\times10^{-11}\mathrm{\,GeV}^{-1}$, that would be detected by the \emph{NuSTAR} instrument are overlaid. The stellar model parameters are described in Tab.~\ref{tab:axion_models}. The spectra are binned to a width of 1 keV, though analysis is performed on unbinned data.}
\label{fig:obs_models}
\end{figure*}

\par In this framework, the differential photon flux per unit energy arriving at Earth is
\begin{equation}
\frac{dN_{\gamma}}{dEdSdt} = \frac{1}{4\pi d^{2}} \frac{d\dot{N_{a}}}{dE}P_{a\gamma}.
\label{eqa:dnr}
\end{equation}
The ALP-photon conversion probability is~\cite{Bassan:2010ya}:
\begin{equation}
P_{a\gamma} = 8.7\times 10^{-6} g_{11}^{2}\left( \dfrac{B_\mathrm{T}}{1~\mu \mathrm{G}}\right)^2 \left( \dfrac{d}{197\,{\rm pc}}\right)^2\dfrac{\sin^2 q}{q^2}
\,\ ,
\label{eqa:prob}
\end{equation}
where $B_\mathrm{T}$ is the transverse magnetic field, namely its component in the plane normal to the path between Earth and Betelgeuse, $q$ is the momentum transfer, and $d$ is the magnetic field length.
\begin{eqnarray}
q &\simeq \left[ 77\,\left(\dfrac{m_{a}}{10^{-10}\,{\rm eV}}\right)^2-0.14\left( \dfrac{n_e}{0.013~{\rm cm}^{-3}}\right) \right] \nonumber \\
& \times \left( \dfrac{d}{197\,{\rm pc}}\right) \left(\dfrac{E}{1\mathrm{\,keV}}\right)^{-1}\, ,
\label{eqa:qform}
\end{eqnarray}
$n_e$ is the electron density \cite{Cordes:2002wz}. Practically, the differential photon flux per unit energy arriving at Earth can be numerically calculated by Eq.~(\ref{eqa:final}) where the parameters are fitted with stellar evolution model and listed in Tab.~\ref{tab:axion_models}. The predicted ALP-photon spectra are illustrated in Fig.~\ref{fig:axion_model_more}.

For small values of $q$ (small ALP mass) and our assumed homogenous ${\bf B}$ field, the expected ALP flux does not depend strongly on the distance to the source, as the drop in flux with distance is compensated by an increase in the conversion probability (Eq.~(\ref{eqa:prob}--\ref{eqa:qform})). Thus we ignore any uncertainty due to the distance to Betelgeuse for the small ALP masses considered here. For large $q$ (for large ALP mass), the increase in the conversion probability with the distance is lost because of incoherence effects.

Even within the simplifying assumption of a uniform regular magnetic field, the exact value of the magnetic field in the direction of Betelgeuse introduces a considerable source of uncertainty in our estimate of the $X$-ray flux. The reported values of the local regular magnetic field, translated to $B_T$ in the direction of Betelgeuse, vary between 0.4 $\mu$G \cite{Jansson2012} and 3.0 $\mu$G~\cite{HarveySmith:2011fe}. Here we are using 1.4 $\mu$G \cite{XuHan2019} as a representative value and 0.4 $\mu$G and 3.0 $\mu$G as the lower and upper bound.

\noindent \textit{Spectral analysis.}---The \emph{NuSTAR} observatory~\cite{Harrison:2013md, Madsen:2015jea, Perez_2019, Roach:2019ctw} is the first focusing high-energy $X$-ray telescope in orbit. Its 3--79 keV energy range is ideal for probing ALP signals from Betelgeuse. \emph{NuSTAR} has two identical co-aligned telescopes, each consisting of an independent optic and focal-plane detector, referred to as FPMA and FPMB. Each telescope subtends a field-of-view of approximately $13^{\prime}\times13^{\prime}$, with a half-power diameter of ${\sim}60^{\prime\prime}$ for a point source near the optical axis. 

\par We use a dedicated \emph{NuSTAR} observation of Betelgeuse taken on 23 August 2019 (ObsID 30501012002). We processed the data using the standard \emph{NuSTAR} data reduction pipeline, \textsc{NuSTARdas} \textsc{v1.8.1} distributed in \textsc{heasoft} \textsc{v6.24}, and the latest calibration package (CALDB.indx20191219). We used the flags \texttt{SAAMODE=OPTIMIZED} and \texttt{TENTACLE=YES} to exclude time intervals with elevated instrument backgrounds, coinciding with the telescope passing through the South Atlantic Anomaly (SAA). After this filtering, the total cleaned exposure was 49.2~ks for FPMA, and 48.4~ks for FPMB.

\par We extract spectra with \textsc{nuproducts}, using a circular source region of radius $60^{\prime\prime}$ around the star's equatorial coordinates (RA 88.79293$^\circ$, Dec. 7.40706$^\circ$) \cite{Hipparcos}. We simultaneously extract the instrument response files used to generate the ALP signal spectra that would be observed in this same region, in particular the Auxiliary Response File (ARF) which defines the energy-dependent effective area for this source region, and the Redistribution Matrix File (RMF) which contains the detector quantum efficiency and energy resolution~\cite{Harrison:2013md}.

\par We extract background spectra from nearby regions, as shown in Fig.~\ref{fig:obs_reg}. These regions are separated from the center of the source region by at least $120^{\prime\prime}$ in order to ensure that contamination from the source region is below the level of  $\mathcal{O}(1\%)$. In addition, we observe a point source near the edge of both FPMs, whose position is consistent with Chandra source $\mathrm{CXO\,J055520.2{+}072002}$ \cite{Evans_2010}. This source is not detected above 10\,keV, but to be conservative we choose the background region to be at least $60^{\prime\prime}$ from this object. The background region is chosen to be on the same detector chip as the source region, in order to properly describe any spatially-varying backgrounds. In particular, the \emph{NuSTAR} instrumental background, including $X$-ray lines resulting from fluorescence/activation of the instrument materials, is known to vary between detector chips, and the stray light from the cosmic $X$-ray background varies radially along the detector plane~\cite{Wik:2014}. Our results are robust to alternate choices of background region, as shown in the Supplementary Material. 

\par The observed $X$-ray spectra for FPMA and FPMB in the source region and background region are shown in top panel of Fig.~\ref{fig:obs_models}. Both source and background rates are higher for FPMB than for FPMA, due to the higher instrumental background in this detector. The background-subtracted source spectra are shown in the lower panel of Fig.~\ref{fig:obs_models}. These were prepared by normalizing the background spectra by the area of the source extraction region, following the procedure in \textsc{nuproducts}. After background subtraction, the source spectrum fluctuates around zero counts. We confirm that our upper limit on the background-subtracted count rate from Betelgeuse is consistent with that measured in soft $X$-rays (0.3--8 keV) using data from \emph{Chandra} (see the Supplementary Material). 

The source spectra after the background subtraction is compared to examples of the predicted $X$-ray spectra from ALP production, for the case of $B_\mathrm{T} = 1.4~\mu \mathrm{G}$, ${m_{a}=10^{-11}\mathrm{\,eV}}$ and $g_{a\gamma}=1.5\times10^{-11}\mathrm{\,GeV}^{-1}$, also shown in the lower panel of Fig.~\ref{fig:obs_models}. We numerically calculate the ALP-produced photon spectrum $dN_{\gamma}/dE_{\gamma}$ that would be detected by \emph{NuSTAR} by folding the predicted spectrum from Eq.~(\ref{eqa:final}) through the instrument response files extracted by \textsc{nuproducts} for this source region. 

\noindent \textit{Data analysis and results.}---Based on the predicted number of photons from ALPs ($N_\mathrm{ax}$) and the expected number of background events ($N_\mathrm{bkg}$) in our source region, we first optimized the energy range used for our analysis by maximizing the figure-of-merit $N_\mathrm{ax}$/$\sqrt{N_\mathrm{bkg}}$. This was done before inspecting our source data. Because of the difference in shape for the predicted ALP spectra, we use three different energy ranges: 10--60 keV for the model with $t_\mathrm{cc}$ = 1.55$\times 10^{5}$ yr, 10--70 keV for $t_\mathrm{cc}$ = 6900--23000 yr, and 10--79 keV for $t_\mathrm{cc}$ = 1.4--3700 yr. The number of source events ($N_\mathrm{obs}$) and expected background events ($N_\mathrm{bkg}$) in the optimized energy ranges for FPMA and FPMB are listed in Tab.~\ref{tab:obs_bkg}. The observed events in Betelgeuse source region are consistent with the expected background events within the statistic uncertainties for both FPMA and FPMB.
\begin{table}[t]
\begin{center}
\renewcommand{\arraystretch}{1.5}
\begin{tabular}{c|>{\centering}m{1.1cm}|>{\centering}m{1.1cm}|>{\centering}m{1.1cm}|c}
    \hline
    \hline
    \multirow{2}{*}{Photon Energy} &\multicolumn{2}{c|}{FPMA} & \multicolumn{2}{c}{FPMB}  \\
    \cline{2-5}
    &$N_\mathrm{obs}$ &$N_\mathrm{bkg}$ &$N_\mathrm{obs}$ &$N_\mathrm{bkg}$ \\
    \hline
    10---60 keV & 313 & 315.8 & 352 & 362.7 \\
    \hline
    10---70 keV & 354 &359.8 & 397 & 406.4 \\
    \hline
    10---79 keV & 384 & 392.7 &433 & 441.2 \\
    \hline
    \hline
\end{tabular}
\end{center}
\caption{Observed events in the source region and expected background events, after normalization to the source region area, for FPMA and FPMB.}
\label{tab:obs_bkg}
\end{table}

To fit the data, an unbinned likelihood function is constructed as~\cite{Junk:1999kv} \\
\begin{equation}
\mathcal{L}=\prod_{i=1}^{n}\mathcal{L}_{i} \times \prod_{i=1}^{n}\mathrm{Gauss}(\delta_\mathrm{bkg}^{i}, \sigma_\mathrm{bkg}^{i}),
\end{equation}
where
\begin{equation}
\begin{split}
\mathcal{L}_{i}=& \mathrm{Poisson}(N_\mathrm{obs}|N_\mathrm{exp}) \\
                &\times\prod_{j=1}^{N_\mathrm{obs}}\left[\frac{N_\mathrm{ax}P_\mathrm{ax}(E_\gamma^{j})}{N_\mathrm{exp}}+\frac{N_\mathrm{bkg}(1+\delta_\mathrm{bkg})P_\mathrm{bkg}(E_\gamma^{j})}{N_\mathrm{exp}}\right]
\end{split}
\end{equation}
Here, $N_\mathrm{obs}$ is the total number of events observed in our source region, and $N_\mathrm{exp} = N_\mathrm{ax}+N_\mathrm{bkg} \cdot (1+\delta_\mathrm{bkg}$) is the total number of events expected in our source region for the case of an ALP signal. $P_\mathrm{ax}(E_\gamma)$ is the energy-dependent ALP signal PDF, defined for given $m_{a}$, $g_{a\gamma}$, $t_\mathrm{cc}$, and $B_{T}$ (examples are shown in the lower panel of Fig.~\ref{fig:obs_models} and Fig.~\ref{fig:axion_model_more}). $P_\mathrm{bkg}(E_\gamma)$ is the background PDF, obtained by normalizing the background spectrum to the source region size using \textsc{nuproducts}, as described above. $\delta_\mathrm{bkg}$ and $\sigma_\mathrm{bkg}$ are the nuisance parameter and fractional systematic uncertainty of the background; Gauss($\delta_\mathrm{bkg}$, $\sigma_\mathrm{bkg}$) is the Gaussian penalty term. Given the statistics of expected background events in the observation region, $\sigma_\mathrm{bkg}$ is set at 10\% for both FPMA and FPMB, but allowed with independent Gaussian fluctuation.

The standard profile likelihood test statistic \cite{Cowan:2010js, Feldman:1997qc} is used to derive constraints on the ALP-photon coupling $g_{a\gamma}$. The test statistic $q$ is defined as
\begin{equation}
q(g_\mathrm{test})=\left\{
\begin{aligned}
&-2\ln\frac{\mathcal{L}_\mathrm{max}(g_\mathrm{test}, \dot{\theta})} {\mathcal{L}_\mathrm{max}(g_\mathrm{best}, \hat{\theta})}, &g_\mathrm{test}\geq g_\mathrm{best} \\
&~0, &g_\mathrm{test}< g_\mathrm{best}
\end{aligned}
\right.
\end{equation}
For each choice of $m_{a}$, $t_\mathrm{cc}$, and $B_{T}$, we scan through the ALP-photon coupling $g_\mathrm{test}$, and perform two maximum likelihood fits, one with the $g_{a\gamma}$ as its best fit value $g_\mathrm{best}$, and the other with $g_{a\gamma}$ fixed at $g_\mathrm{test}$. The nuisance parameters are all allowed to vary in both to achieve the best fit. We derive the 95\% C.L. upper limit on $g_{a\gamma}$ assuming $q(g_\mathrm{test}$) follows a half-$\chi^{2}$ distribution with a single degree of freedom \cite{Cowan:2010js}.

The main sources of uncertainty on the ALP signal are the choice of stellar model ($t_\mathrm{cc}$) and assumed value of the transverse local Galactic magnetic field strength ($B_\mathrm{T}$). Rather than accounting for these as nuisance parameters in the likelihood function, we derive separate 95\% C.L. limits on $g_{a\gamma}$ for each of the 13 stellar models for each of our three assumed values of $B_\mathrm{T}$. 

\begin{figure}[t]
\centering
\includegraphics[width=\columnwidth]{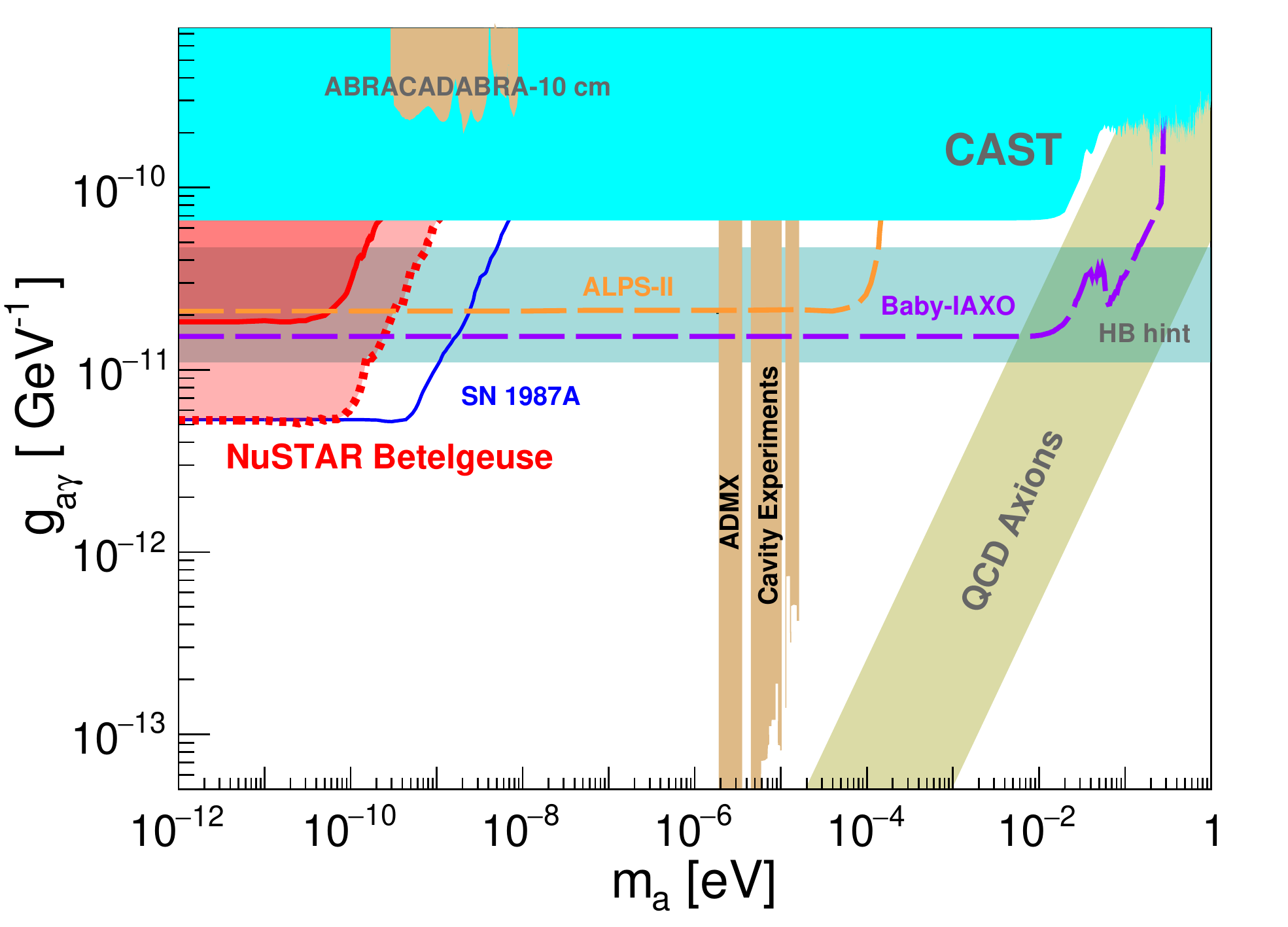}
\caption{Comparison of 95\% upper limits of $g_{a\gamma}$ from \emph{NuSTAR} Betelgeuse observation is shown in red band: upper solid line for the most conservative stellar model and $B_\mathrm{T}$, lower dashed line for the most optimistic stellar model and $B_\mathrm{T}$, and the interval for the other combinations. Overlaid are the region of interest for QCD axion \cite{DiLuzio:2016sbl}, the current limits from CAST \cite{cast:2017ftl}, ADMX \cite{admx:2019fqb}, ABRACADABRA-10cm \cite{Ouellet:2018beu} and SN 1987A~\cite{payez:2014xsa}, the projected sensitivity of ALPS-II \cite{alps2:2013ywa} and Baby-IAXO \cite{Armengaud:2019uso, DiVecchia:2019ejf}. Also labelled are the ALP regions suggested by hints of excess gamma-ray transparency or cooling of horizontal-branch stars~\cite{Giannotti:2015kwo,Giannotti:2017hny}.
}
\label{fig:upperlimts}
\end{figure}

The final 95\% C.L. upper limit on $g_{a\gamma}$ is shown in Fig.~\ref{fig:upperlimts} and Fig.~\ref{fig:upperlimts_lowmass}. The region labeled in red is excluded by this work while the width of the light red band reflects the uncertainty due to choice of stellar model and $B_\mathrm{T}$. Using our most conservative assumptions ($t_\mathrm{cc}$ = 1.55$\times 10^{5}$ yr and $B_\mathrm{T}$ = 0.4 $\mu G$), we set an upper limit of $g_{a\gamma}<1.8\times10^{-11}\mathrm{\,GeV}^{-1}$ for $m_{a}<3.5\times10^{-11}\mathrm{\,eV}$. In the scenario that predicts the highest ALP flux ($t_\mathrm{cc}$ = 3.6 yr and $B_\mathrm{T}$ = 3.0 $\mu G$), we derive an upper limit of $g_{a\gamma}<5.2\times10^{-12}\mathrm{\,GeV}^{-1}$ for $m_{a}<5.5\times10^{-11}\mathrm{\,eV}$. 

\begin{figure}[htb!]
\centering
\includegraphics[width=\columnwidth]{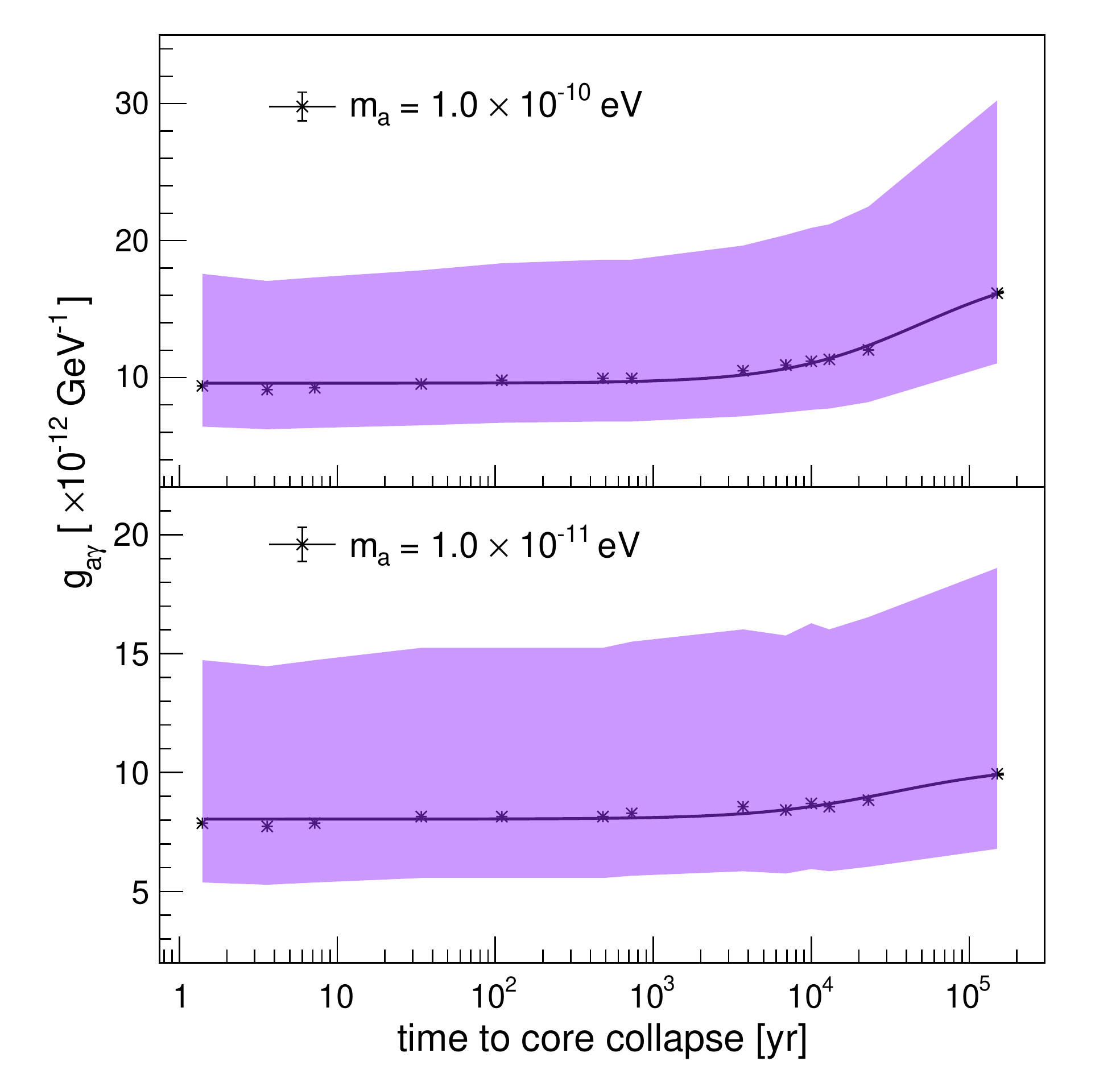}
\caption{Evolution on the derived 95\% C.L. upper limit of $g_{a\gamma}$ with remaining time until the core-collapse for Betelgeuse. Results are shown for $m_{a}=1.0\times10^{-11} ~\textrm{eV}$ and $m_{a}=1.0\times10^{-10} ~\textrm{eV}$ assuming $B_{T}$ = 1.4 $\mu G$ (more masses are shown in the Supplementary Material). The solid black line shows the fitting. The width of the violet band indicates the uncertainty due to choice of $B_{T}$.}
\label{fig:gar_evolution}
\end{figure}

The uncertainty in our derived limit is dominated by our choice of $B_\mathrm{T}$, since the ALP-photon conversion probability of Eq.~(\ref{eqa:prob}) scales as $B_\mathrm{T}^{2}$. The separate contributions to the uncertainty are illustrated in Fig.~\ref{fig:gar_evolution}, which shows the evolution of our derived $g_{a\gamma}$ for different $t_\mathrm{cc}$. The solid black line is a fit to the $g_{a\gamma}$ derived for $B_\mathrm{T}$ = 1.4 $\mu G$, shown by the black points. The width of the magenta band then indicates the uncertainty due to our lower and upper bounds on $B_\mathrm{T}$. The dependence on our assumed $B_\mathrm{T}$ is further illustrated in Fig.~\ref{fig:results_garbt}, where we show our  constraint in terms of $GeV^{-1} \sqrt{\mu G}$ for our range of $t_{cc}$.

\par \noindent \textit{Conclusions and Discussions.}---Fig.~\ref{fig:upperlimts} shows the current theoretical and experimental ALP landscape. This includes the current constraints from CAST~\cite{cast:2017ftl}, ABRACADABRA-10cm~\cite{Ouellet:2018beu}, and cavity experiments such as ADMX~\cite{admx:2019fqb}. The yellow band indicates the preferred region in which the axion couples to quantum chromodynamics (QCD) and could be a solution to the strong-CP problem \cite{DiLuzio:2016sbl}, and the hatched band indicates the region where an ALP could explain the observed excess cooling of horizontal branch stars~\cite{Giannotti:2015kwo,Giannotti:2017hny}. The limits derived in this work exceed those set by CAST by a factor of $\sim$3 for $m_{a}<3.5\times10^{-11}\mathrm{\,eV}$, and are comparable in this low-mass space to the sensitivity of even next-generation experiments such as ALPS-II~\cite{alps2:2013ywa} and BabyIAXO~\cite{Armengaud:2019uso, DiVecchia:2019ejf}.

Fig.~\ref{fig:upperlimts_lowmass} compares our result to other astrophysical constraints in the low-mass ALP regime. Our most conservative limit exceeds the bounds from extra-galactic supernovae \cite{Meyer:2020vzy} for $m_{a} < 10^{-10}$ eV, and in our most optimistic ALP flux scenario our limit is comparable to that derived from the non-observation of gamma-rays from SN 1987A~\cite{payez:2014xsa}. However, the supernova bounds have been questioned due to the modeling of core-collapse supernovae. For $m_{a} < 10^{-11}$ eV, our limits are superseded by those derived from the lack of spectral variation observed by Chandra in the active galactic nucleus NGC 1275 \cite{chandra:2019uqt} and in the core of M87 \cite{m87:2017yvc}. We caution, though, that more recent analysis shows that these results could be weakened by several orders of magnitude depending on the relative magnitude of the regular and turbulent intracluster magnetic fields that are assumed \cite{Libanov:2019fzq}.

We emphasize that since each of these astrophysical constraints has unique sources of systematic error that may affect the final result, it is worthwhile to survey similar regions of parameter space with multiple techniques. 
The constraints presented here assume conversion in a homogeneous regular magnetic field, which allows for consistent comparison with previous astrophysical bounds, e.g. SN 1987A~\cite{payez:2014xsa}.
Although the magnetic structure of the Galaxy is certainly more complex, compared to more distant sources, the proximity of Betelgeuse should allow future observations to better constrain the relevant small-scale variations in magnetic structure, especially the ultra-local magnetic fields of the Solar region which are not yet completely mapped (e.g.~\cite{Salvati2010,Frisch2012,XuHan2019}).
The combination of the novel ALP constraint presented here with multiple overlapping astrophysical constraints, each with separate modeling assumptions and uncertainties, builds confidence in the robustness of the exclusion of this corner of parameter space for low-mass ALPs.

We finally comment that a synergy between our astrophysical approach and direct ALP searches might lead to surprises and also unexpected benefits. Indeed, it might be that a future low-mass ALP experiment such as ABRACADABRA \cite{Ouellet:2019tlz}, DM-Radio \cite{dm_radio}, or IAXO \cite{Armengaud:2019uso} would discover an ALP in the region where optimistic assumptions on Galactic $B$-field and stellar model would have led to an exclusion from Betelgeuse. In this case one would come back to our original assumptions. Taking into account the typical uncertainty on the $B$-field one would give a lower limit on the time until the core-collapse for Betelgeuse, as shown in Fig.~\ref{fig:gar_evolution} for an ALP mass and a coupling in the range $g_{a\gamma}= (5-30)\times10^{-12}$ $\mathrm{GeV^{-1}}$. Intriguingly ALPs might represent the only possibility to extract such information about the fate of Betelgeuse.

\section{Acknowledgements}
We thank D. R. Wik and S. Rossland for helpful discussions on the \emph{NuSTAR} instrument background;  J.L. Han and S. Zhang for discussions concerning the Galactic magnetic field in the direction of Betelgeuse; Benjamin Safdi, Christopher Dessert, John Beacom, Shunsaku Horiuchi and Kenny C. Y. Ng for helpful comments and discussions. The \emph{NuSTAR} observations described in this work were awarded under NASA Grant No. 80NSSC20K0031. The computational aspects of this work made extensive use of the following packages: \textsc{saoimage ds9} distributed by the Smithsonian Astrophysical Observatory; the \textsc{scipy} ecosystem, particularly \textsc{matplotlib} and \textsc{numpy}; and \textsc{astropy}, a community-developed core \textsc{python} package for Astronomy. This research has made use of data and software provided by the High Energy Astrophysics Science Archive Research Center (HEASARC), which is a service of the Astrophysics Science Division at NASA/GSFC and the High Energy Astrophysics Division of the Smithsonian Astrophysical Observatory. We thank the \emph{NuSTAR} Operations, Software and Calibration teams for support with the execution and analysis of these observations. This research made use of the \emph{NuSTAR} Data Analysis Software (NuSTARDAS), jointly developed by the ASI Science Data Center (ASDC, Italy) and the California Institute of Technology (USA). M.X. and M.G. were supported by NASA Grant No. 80NSSC20K0031. O.S. has been supported by the Agenzia Spaziale Italiana (ASI) and the Instituto Nazionale di Astrofsica (INAF) under the agreement  n.   2017-14-H.0  - attivit\`a  di  studio  per  la  comunit\`a  scientifica  di  Astrofisica  delle  Alte Energie  e  Fisica Astroparticellare. A.M. is partially supported by the Italian Istituto Nazionale di Fisica Nucleare (INFN) through the ``Theoretical Astroparticle Physics'' project and by the research grant number 2017W4HA7S ``NAT-NET: Neutrino and Astroparticle Theory Network'' under the program PRIN 2017 funded by the Italian Ministero dell'Universit\`a e della Ricerca (MUR). B.G. was supported under NASA contract NNG08FD60C.

\bibliography{Betelgeuse}

\begin{thebibliography}{62}%
\makeatletter
\providecommand \@ifxundefined [1]{%
 \@ifx{#1\undefined}
}%
\providecommand \@ifnum [1]{%
 \ifnum #1\expandafter \@firstoftwo
 \else \expandafter \@secondoftwo
 \fi
}%
\providecommand \@ifx [1]{%
 \ifx #1\expandafter \@firstoftwo
 \else \expandafter \@secondoftwo
 \fi
}%
\providecommand \natexlab [1]{#1}%
\providecommand \enquote  [1]{``#1''}%
\providecommand \bibnamefont  [1]{#1}%
\providecommand \bibfnamefont [1]{#1}%
\providecommand \citenamefont [1]{#1}%
\providecommand \href@noop [0]{\@secondoftwo}%
\providecommand \href [0]{\begingroup \@sanitize@url \@href}%
\providecommand \@href[1]{\@@startlink{#1}\@@href}%
\providecommand \@@href[1]{\endgroup#1\@@endlink}%
\providecommand \@sanitize@url [0]{\catcode `\\12\catcode `\$12\catcode
  `\&12\catcode `\#12\catcode `\^12\catcode `\_12\catcode `\%12\relax}%
\providecommand \@@startlink[1]{}%
\providecommand \@@endlink[0]{}%
\providecommand \url  [0]{\begingroup\@sanitize@url \@url }%
\providecommand \@url [1]{\endgroup\@href {#1}{\urlprefix }}%
\providecommand \urlprefix  [0]{URL }%
\providecommand \Eprint [0]{\href }%
\providecommand \doibase [0]{http://dx.doi.org/}%
\providecommand \selectlanguage [0]{\@gobble}%
\providecommand \bibinfo  [0]{\@secondoftwo}%
\providecommand \bibfield  [0]{\@secondoftwo}%
\providecommand \translation [1]{[#1]}%
\providecommand \BibitemOpen [0]{}%
\providecommand \bibitemStop [0]{}%
\providecommand \bibitemNoStop [0]{.\EOS\space}%
\providecommand \EOS [0]{\spacefactor3000\relax}%
\providecommand \BibitemShut  [1]{\csname bibitem#1\endcsname}%
\let\auto@bib@innerbib\@empty
\bibitem [{\citenamefont {Di~Luzio}\ \emph {et~al.}(2020)\citenamefont
  {Di~Luzio}, \citenamefont {Giannotti}, \citenamefont {Nardi},\ and\
  \citenamefont {Visinelli}}]{DiLuzio:2020wdo}%
  \BibitemOpen
  \bibfield  {author} {\bibinfo {author} {\bibfnamefont {Luca}\ \bibnamefont
  {Di~Luzio}}, \bibinfo {author} {\bibfnamefont {Maurizio}\ \bibnamefont
  {Giannotti}}, \bibinfo {author} {\bibfnamefont {Enrico}\ \bibnamefont
  {Nardi}}, \ and\ \bibinfo {author} {\bibfnamefont {Luca}\ \bibnamefont
  {Visinelli}},\ }\bibfield  {title} {\enquote {\bibinfo {title} {{The
  landscape of QCD axion models}},}\ }\href {\doibase
  10.1016/j.physrep.2020.06.002} {\bibfield  {journal} {\bibinfo  {journal}
  {Phys. Rept.}\ }\textbf {\bibinfo {volume} {870}},\ \bibinfo {pages} {1--117}
  (\bibinfo {year} {2020})},\ \Eprint {http://arxiv.org/abs/2003.01100}
  {arXiv:2003.01100 [hep-ph]} \BibitemShut {NoStop}%
\bibitem [{\citenamefont {Jaeckel}\ and\ \citenamefont
  {Ringwald}(2010)}]{Jaeckel:2010ni}%
  \BibitemOpen
  \bibfield  {author} {\bibinfo {author} {\bibfnamefont {Joerg}\ \bibnamefont
  {Jaeckel}}\ and\ \bibinfo {author} {\bibfnamefont {Andreas}\ \bibnamefont
  {Ringwald}},\ }\bibfield  {title} {\enquote {\bibinfo {title} {{The
  Low-Energy Frontier of Particle Physics}},}\ }\href {\doibase
  10.1146/annurev.nucl.012809.104433} {\bibfield  {journal} {\bibinfo
  {journal} {Ann. Rev. Nucl. Part. Sci.}\ }\textbf {\bibinfo {volume} {60}},\
  \bibinfo {pages} {405--437} (\bibinfo {year} {2010})},\ \Eprint
  {http://arxiv.org/abs/1002.0329} {arXiv:1002.0329 [hep-ph]} \BibitemShut
  {NoStop}%
\bibitem [{\citenamefont {Raffelt}\ and\ \citenamefont
  {Stodolsky}(1988)}]{Raffelt:1987im}%
  \BibitemOpen
  \bibfield  {author} {\bibinfo {author} {\bibfnamefont {Georg}\ \bibnamefont
  {Raffelt}}\ and\ \bibinfo {author} {\bibfnamefont {Leo}\ \bibnamefont
  {Stodolsky}},\ }\bibfield  {title} {\enquote {\bibinfo {title} {{Mixing of
  the Photon with Low Mass Particles}},}\ }\href {\doibase
  10.1103/PhysRevD.37.1237} {\bibfield  {journal} {\bibinfo  {journal} {Phys.
  Rev. D}\ }\textbf {\bibinfo {volume} {37}},\ \bibinfo {pages} {1237}
  (\bibinfo {year} {1988})}\BibitemShut {NoStop}%
\bibitem [{\citenamefont {{Sikivie}}(2020)}]{Sikivie:2020zpn}%
  \BibitemOpen
  \bibfield  {author} {\bibinfo {author} {\bibfnamefont {Pierre}\ \bibnamefont
  {{Sikivie}}},\ }\bibfield  {title} {\enquote {\bibinfo {title} {{Invisible
  Axion Search Methods}},}\ }\href@noop {} {\bibfield  {journal} {\bibinfo
  {journal} {arXiv e-prints}\ } (\bibinfo {year} {2020})},\ \Eprint
  {http://arxiv.org/abs/2003.02206} {arXiv:2003.02206 [hep-ph]} \BibitemShut
  {NoStop}%
\bibitem [{\citenamefont {Irastorza}\ and\ \citenamefont
  {Redondo}(2018)}]{Irastorza:2018dyq}%
  \BibitemOpen
  \bibfield  {author} {\bibinfo {author} {\bibfnamefont {Igor~G.}\ \bibnamefont
  {Irastorza}}\ and\ \bibinfo {author} {\bibfnamefont {Javier}\ \bibnamefont
  {Redondo}},\ }\bibfield  {title} {\enquote {\bibinfo {title} {{New
  Experimental Approaches in the Search for Axion-Like Particles}},}\ }\href
  {\doibase 10.1016/j.ppnp.2018.05.003} {\bibfield  {journal} {\bibinfo
  {journal} {Prog. Part. Nucl. Phys.}\ }\textbf {\bibinfo {volume} {102}},\
  \bibinfo {pages} {89--159} (\bibinfo {year} {2018})},\ \Eprint
  {http://arxiv.org/abs/1801.08127} {arXiv:1801.08127 [hep-ph]} \BibitemShut
  {NoStop}%
\bibitem [{\citenamefont {Raffelt}(1986)}]{Raffelt:1985nk}%
  \BibitemOpen
  \bibfield  {author} {\bibinfo {author} {\bibfnamefont {Georg~G.}\
  \bibnamefont {Raffelt}},\ }\bibfield  {title} {\enquote {\bibinfo {title}
  {{Astrophysical Axion Bounds Diminished by Screening Effects}},}\ }\href
  {\doibase 10.1103/PhysRevD.33.897} {\bibfield  {journal} {\bibinfo  {journal}
  {Phys. Rev. D}\ }\textbf {\bibinfo {volume} {33}},\ \bibinfo {pages} {897}
  (\bibinfo {year} {1986})}\BibitemShut {NoStop}%
\bibitem [{\citenamefont {Ayala}\ \emph {et~al.}(2014)\citenamefont {Ayala},
  \citenamefont {Domínguez}, \citenamefont {Giannotti}, \citenamefont
  {Mirizzi},\ and\ \citenamefont {Straniero}}]{Ayala:2014pea}%
  \BibitemOpen
  \bibfield  {author} {\bibinfo {author} {\bibfnamefont {Adrian}\ \bibnamefont
  {Ayala}}, \bibinfo {author} {\bibfnamefont {Inma}\ \bibnamefont
  {Domínguez}}, \bibinfo {author} {\bibfnamefont {Maurizio}\ \bibnamefont
  {Giannotti}}, \bibinfo {author} {\bibfnamefont {Alessandro}\ \bibnamefont
  {Mirizzi}}, \ and\ \bibinfo {author} {\bibfnamefont {Oscar}\ \bibnamefont
  {Straniero}},\ }\bibfield  {title} {\enquote {\bibinfo {title} {{Revisiting
  the Bound on Axion-Photon Coupling from Globular Clusters}},}\ }\href
  {\doibase 10.1103/PhysRevLett.113.191302} {\bibfield  {journal} {\bibinfo
  {journal} {Phys. Rev. Lett.}\ }\textbf {\bibinfo {volume} {113}},\ \bibinfo
  {pages} {191302} (\bibinfo {year} {2014})},\ \Eprint
  {http://arxiv.org/abs/1406.6053} {arXiv:1406.6053 [astro-ph.SR]} \BibitemShut
  {NoStop}%
\bibitem [{\citenamefont {Straniero}\ \emph {et~al.}(2015)\citenamefont
  {Straniero}, \citenamefont {Ayala}, \citenamefont {Giannotti}, \citenamefont
  {Mirizzi},\ and\ \citenamefont {Dominguez}}]{Straniero:2015nvc}%
  \BibitemOpen
  \bibfield  {author} {\bibinfo {author} {\bibfnamefont {Oscar}\ \bibnamefont
  {Straniero}}, \bibinfo {author} {\bibfnamefont {Adrian}\ \bibnamefont
  {Ayala}}, \bibinfo {author} {\bibfnamefont {Maurizio}\ \bibnamefont
  {Giannotti}}, \bibinfo {author} {\bibfnamefont {Alessandro}\ \bibnamefont
  {Mirizzi}}, \ and\ \bibinfo {author} {\bibfnamefont {Inma}\ \bibnamefont
  {Dominguez}},\ }\bibfield  {title} {\enquote {\bibinfo {title} {{Axion-Photon
  Coupling: Astrophysical Constraints}},}\ }in\ \href {\doibase
  10.3204/DESY-PROC-2015-02/straniero\_oscar} {\emph {\bibinfo {booktitle}
  {{11th Patras Workshop on Axions, WIMPs and WISPs}}}}\ (\bibinfo {year}
  {2015})\ pp.\ \bibinfo {pages} {77--81}\BibitemShut {NoStop}%
\bibitem [{\citenamefont {Carenza}\ \emph {et~al.}(2020)\citenamefont
  {Carenza}, \citenamefont {Straniero}, \citenamefont {D\"obrich},
  \citenamefont {Giannotti}, \citenamefont {Lucente},\ and\ \citenamefont
  {Mirizzi}}]{Carenza:2020zil}%
  \BibitemOpen
  \bibfield  {author} {\bibinfo {author} {\bibfnamefont {Pierluca}\
  \bibnamefont {Carenza}}, \bibinfo {author} {\bibfnamefont {Oscar}\
  \bibnamefont {Straniero}}, \bibinfo {author} {\bibfnamefont {Babette}\
  \bibnamefont {D\"obrich}}, \bibinfo {author} {\bibfnamefont {Maurizio}\
  \bibnamefont {Giannotti}}, \bibinfo {author} {\bibfnamefont {Giuseppe}\
  \bibnamefont {Lucente}}, \ and\ \bibinfo {author} {\bibfnamefont
  {Alessandro}\ \bibnamefont {Mirizzi}},\ }\bibfield  {title} {\enquote
  {\bibinfo {title} {{Constraints on the coupling with photons of heavy
  axion-like-particles from Globular Clusters}},}\ }\href {\doibase
  10.1016/j.physletb.2020.135709} {\bibfield  {journal} {\bibinfo  {journal}
  {Phys. Lett. B}\ }\textbf {\bibinfo {volume} {809}},\ \bibinfo {pages}
  {135709} (\bibinfo {year} {2020})},\ \Eprint
  {http://arxiv.org/abs/2004.08399} {arXiv:2004.08399 [hep-ph]} \BibitemShut
  {NoStop}%
\bibitem [{\citenamefont {Anastassopoulos}\ \emph {et~al.}(2017)\citenamefont
  {Anastassopoulos} \emph {et~al.}}]{cast:2017ftl}%
  \BibitemOpen
  \bibfield  {author} {\bibinfo {author} {\bibfnamefont {V.}~\bibnamefont
  {Anastassopoulos}} \emph {et~al.} (\bibinfo {collaboration} {CAST}),\
  }\bibfield  {title} {\enquote {\bibinfo {title} {{New CAST Limit on the
  Axion-Photon Interaction}},}\ }\href {\doibase 10.1038/nphys4109} {\bibfield
  {journal} {\bibinfo  {journal} {Nature Phys.}\ }\textbf {\bibinfo {volume}
  {13}},\ \bibinfo {pages} {584--590} (\bibinfo {year} {2017})},\ \Eprint
  {http://arxiv.org/abs/1705.02290} {arXiv:1705.02290 [hep-ex]} \BibitemShut
  {NoStop}%
\bibitem [{\citenamefont {Payez}\ \emph {et~al.}(2015)\citenamefont {Payez},
  \citenamefont {Evoli}, \citenamefont {Fischer}, \citenamefont {Giannotti},
  \citenamefont {Mirizzi},\ and\ \citenamefont {Ringwald}}]{payez:2014xsa}%
  \BibitemOpen
  \bibfield  {author} {\bibinfo {author} {\bibfnamefont {Alexandre}\
  \bibnamefont {Payez}}, \bibinfo {author} {\bibfnamefont {Carmelo}\
  \bibnamefont {Evoli}}, \bibinfo {author} {\bibfnamefont {Tobias}\
  \bibnamefont {Fischer}}, \bibinfo {author} {\bibfnamefont {Maurizio}\
  \bibnamefont {Giannotti}}, \bibinfo {author} {\bibfnamefont {Alessandro}\
  \bibnamefont {Mirizzi}}, \ and\ \bibinfo {author} {\bibfnamefont {Andreas}\
  \bibnamefont {Ringwald}},\ }\bibfield  {title} {\enquote {\bibinfo {title}
  {{Revisiting the SN1987A Gamma-Ray Limit on Ulltralight Axion-Like
  Particles}},}\ }\href {\doibase 10.1088/1475-7516/2015/02/006} {\bibfield
  {journal} {\bibinfo  {journal} {J. Cosm}\ }\textbf {\bibinfo {volume} {02}},\
  \bibinfo {pages} {006} (\bibinfo {year} {2015})},\ \Eprint
  {http://arxiv.org/abs/1410.3747} {arXiv:1410.3747 [astro-ph.HE]} \BibitemShut
  {NoStop}%
\bibitem [{\citenamefont {Meyer}\ and\ \citenamefont
  {Petrushevska}(2020)}]{Meyer:2020vzy}%
  \BibitemOpen
  \bibfield  {author} {\bibinfo {author} {\bibfnamefont {Manuel}\ \bibnamefont
  {Meyer}}\ and\ \bibinfo {author} {\bibfnamefont {Tanja}\ \bibnamefont
  {Petrushevska}},\ }\bibfield  {title} {\enquote {\bibinfo {title} {{Search
  for Axionlike-Particle-Induced Prompt Gamma-Ray Emission from Extragalactic
  Core-Collapse Supernovae with the Fermi Large Area Telescope}},}\ }\href
  {\doibase 10.1103/PhysRevLett.124.231101} {\bibfield  {journal} {\bibinfo
  {journal} {Phys. Rev. Lett.}\ }\textbf {\bibinfo {volume} {124}},\ \bibinfo
  {pages} {231101} (\bibinfo {year} {2020})},\ \Eprint
  {http://arxiv.org/abs/2006.06722v2} {arXiv:2006.06722v2 [astro-ph.HE]}
  \BibitemShut {NoStop}%
\bibitem [{\citenamefont {{Forrest}}\ \emph {et~al.}(1980)\citenamefont
  {{Forrest}} \emph {et~al.}}]{Forrest:1980}%
  \BibitemOpen
  \bibfield  {author} {\bibinfo {author} {\bibfnamefont {D.~J.}\ \bibnamefont
  {{Forrest}}} \emph {et~al.},\ }\bibfield  {title} {\enquote {\bibinfo {title}
  {{The Gamma Ray Spectrometer for the Solar Maximum Mission}},}\ }\href
  {\doibase 10.1007/BF00151381} {\bibfield  {journal} {\bibinfo  {journal}
  {Solar Physics}\ }\textbf {\bibinfo {volume} {65}},\ \bibinfo {pages}
  {15--23} (\bibinfo {year} {1980})}\BibitemShut {NoStop}%
\bibitem [{\citenamefont {Marsh}\ \emph {et~al.}(2017)\citenamefont {Marsh},
  \citenamefont {Russell}, \citenamefont {Fabian}, \citenamefont {McNamara},
  \citenamefont {Nulsen},\ and\ \citenamefont {Reynolds}}]{m87:2017yvc}%
  \BibitemOpen
  \bibfield  {author} {\bibinfo {author} {\bibfnamefont {M.C.~David}\
  \bibnamefont {Marsh}}, \bibinfo {author} {\bibfnamefont {Helen~R.}\
  \bibnamefont {Russell}}, \bibinfo {author} {\bibfnamefont {Andrew~C.}\
  \bibnamefont {Fabian}}, \bibinfo {author} {\bibfnamefont {Brian~P.}\
  \bibnamefont {McNamara}}, \bibinfo {author} {\bibfnamefont {Paul}\
  \bibnamefont {Nulsen}}, \ and\ \bibinfo {author} {\bibfnamefont
  {Christopher~S.}\ \bibnamefont {Reynolds}},\ }\bibfield  {title} {\enquote
  {\bibinfo {title} {{A New Bound on Axion-Like Particles}},}\ }\href {\doibase
  10.1088/1475-7516/2017/12/036} {\bibfield  {journal} {\bibinfo  {journal} {J.
  Cosmol. Astropart. Phys.}\ }\textbf {\bibinfo {volume} {12}},\ \bibinfo
  {pages} {036} (\bibinfo {year} {2017})},\ \Eprint
  {http://arxiv.org/abs/1703.07354} {arXiv:1703.07354 [hep-ph]} \BibitemShut
  {NoStop}%
\bibitem [{\citenamefont {Reynolds}\ \emph {et~al.}(2019)\citenamefont
  {Reynolds}, \citenamefont {Marsh}, \citenamefont {Russell}, \citenamefont
  {Fabian}, \citenamefont {Smith}, \citenamefont {Tombesi},\ and\ \citenamefont
  {Veilleux}}]{chandra:2019uqt}%
  \BibitemOpen
  \bibfield  {author} {\bibinfo {author} {\bibfnamefont {Christopher~S.}\
  \bibnamefont {Reynolds}}, \bibinfo {author} {\bibfnamefont {M.~C.~David}\
  \bibnamefont {Marsh}}, \bibinfo {author} {\bibfnamefont {Helen~R.}\
  \bibnamefont {Russell}}, \bibinfo {author} {\bibfnamefont {Andrew~C.}\
  \bibnamefont {Fabian}}, \bibinfo {author} {\bibfnamefont {Robyn}\
  \bibnamefont {Smith}}, \bibinfo {author} {\bibfnamefont {Francesco}\
  \bibnamefont {Tombesi}}, \ and\ \bibinfo {author} {\bibfnamefont {Sylvain}\
  \bibnamefont {Veilleux}},\ }\bibfield  {title} {\enquote {\bibinfo {title}
  {{Astrophysical Limits on Very Light Axion-Like Particles from Chandra
  Grating Spectroscopy of NGC 1275}},}\ }\href {\doibase
  10.3847/1538-4357/ab6a0c} {\bibfield  {journal} {\bibinfo  {journal}
  {Astrophys. J.}\ }\textbf {\bibinfo {volume} {890}},\ \bibinfo {pages} {59}
  (\bibinfo {year} {2019})},\ \Eprint {http://arxiv.org/abs/1907.05475}
  {arXiv:1907.05475 [hep-ph]} \BibitemShut {NoStop}%
\bibitem [{\citenamefont {Libanov}\ and\ \citenamefont
  {Troitsky}(2020)}]{Libanov:2019fzq}%
  \BibitemOpen
  \bibfield  {author} {\bibinfo {author} {\bibfnamefont {Maxim}\ \bibnamefont
  {Libanov}}\ and\ \bibinfo {author} {\bibfnamefont {Sergey}\ \bibnamefont
  {Troitsky}},\ }\bibfield  {title} {\enquote {\bibinfo {title} {{On the Impact
  of Magnetic-Field Models in Galaxy Clusters on Constraints on Axion-Like
  Particles from the Lack of Irregularities in High-Energy Spectra of
  Astrophysical Sources}},}\ }\href {\doibase 10.1016/j.physletb.2020.135252}
  {\bibfield  {journal} {\bibinfo  {journal} {Phys. Lett. B}\ }\textbf
  {\bibinfo {volume} {802}},\ \bibinfo {pages} {135252} (\bibinfo {year}
  {2020})},\ \Eprint {http://arxiv.org/abs/1908.03084} {arXiv:1908.03084
  [astro-ph.HE]} \BibitemShut {NoStop}%
\bibitem [{\citenamefont {Carlson}(1995)}]{Carlson:1995wa}%
  \BibitemOpen
  \bibfield  {author} {\bibinfo {author} {\bibfnamefont {E.~D.}\ \bibnamefont
  {Carlson}},\ }\bibfield  {title} {\enquote {\bibinfo {title} {{Pseudoscalar
  Conversion and X-Rays from Stars}},}\ }\href {\doibase
  10.1016/0370-2693(94)01529-L} {\bibfield  {journal} {\bibinfo  {journal}
  {Phys. Lett. B}\ }\textbf {\bibinfo {volume} {344}},\ \bibinfo {pages}
  {245--251} (\bibinfo {year} {1995})}\BibitemShut {NoStop}%
\bibitem [{\citenamefont {{Posson-Brown}}\ \emph {et~al.}(2006)\citenamefont
  {{Posson-Brown}}, \citenamefont {{Kashyap}}, \citenamefont {{Pease}},\ and\
  \citenamefont {{Drake}}}]{PossonBrown:2006zr}%
  \BibitemOpen
  \bibfield  {author} {\bibinfo {author} {\bibfnamefont {Jennifer}\
  \bibnamefont {{Posson-Brown}}}, \bibinfo {author} {\bibfnamefont {Vinay~L.}\
  \bibnamefont {{Kashyap}}}, \bibinfo {author} {\bibfnamefont {Deron~O.}\
  \bibnamefont {{Pease}}}, \ and\ \bibinfo {author} {\bibfnamefont {Jeremy~J.}\
  \bibnamefont {{Drake}}},\ }\bibfield  {title} {\enquote {\bibinfo {title}
  {{Invisible Giant: Chandra's Limits on X-rays from Betelgeuse}},}\
  }\href@noop {} {\bibfield  {journal} {\bibinfo  {journal} {arXiv e-prints}\ }
  (\bibinfo {year} {2006})},\ \Eprint {http://arxiv.org/abs/astro-ph/0606387}
  {arXiv:astro-ph/0606387 [astro-ph]} \BibitemShut {NoStop}%
\bibitem [{\citenamefont {Dessert}\ \emph {et~al.}(2020)\citenamefont
  {Dessert}, \citenamefont {Foster},\ and\ \citenamefont {Safdi}}]{ben2020}%
  \BibitemOpen
  \bibfield  {author} {\bibinfo {author} {\bibfnamefont {Christopher}\
  \bibnamefont {Dessert}}, \bibinfo {author} {\bibfnamefont {Joshua~W.}\
  \bibnamefont {Foster}}, \ and\ \bibinfo {author} {\bibfnamefont
  {Benjamin~R.}\ \bibnamefont {Safdi}},\ }\bibfield  {title} {\enquote
  {\bibinfo {title} {{X-ray Searches for Axions from Super Star Clusters}},}\
  }\href@noop {} {\bibfield  {journal} {\bibinfo  {journal} {arXiv e-prints}\ }
  (\bibinfo {year} {2020})},\ \Eprint {http://arxiv.org/abs/2008.03305}
  {arXiv:2008.03305 [hep-ph]} \BibitemShut {NoStop}%
\bibitem [{\citenamefont {Harper}\ \emph {et~al.}(2008)\citenamefont {Harper},
  \citenamefont {Brown},\ and\ \citenamefont {Guinan}}]{Harper_2008}%
  \BibitemOpen
  \bibfield  {author} {\bibinfo {author} {\bibfnamefont {Graham~M.}\
  \bibnamefont {Harper}}, \bibinfo {author} {\bibfnamefont {Alexander}\
  \bibnamefont {Brown}}, \ and\ \bibinfo {author} {\bibfnamefont {Edward~F.}\
  \bibnamefont {Guinan}},\ }\bibfield  {title} {\enquote {\bibinfo {title} {{A
  New VLA-Hipparcos Distance to Betelgeuse and its Implications}},}\ }\href
  {\doibase 10.1088/0004-6256/135/4/1430} {\bibfield  {journal} {\bibinfo
  {journal} {Astron. J.}\ }\textbf {\bibinfo {volume} {135}},\ \bibinfo {pages}
  {1430--1440} (\bibinfo {year} {2008})}\BibitemShut {NoStop}%
\bibitem [{\citenamefont {Harper}\ \emph {et~al.}(2017)\citenamefont {Harper},
  \citenamefont {Brown}, \citenamefont {Guinan}, \citenamefont {O'Gorman},
  \citenamefont {Richards}, \citenamefont {Kervella},\ and\ \citenamefont
  {Decin}}]{Harper_2017}%
  \BibitemOpen
  \bibfield  {author} {\bibinfo {author} {\bibfnamefont {G.~M.}\ \bibnamefont
  {Harper}}, \bibinfo {author} {\bibfnamefont {A.}~\bibnamefont {Brown}},
  \bibinfo {author} {\bibfnamefont {E.~F.}\ \bibnamefont {Guinan}}, \bibinfo
  {author} {\bibfnamefont {E.}~\bibnamefont {O'Gorman}}, \bibinfo {author}
  {\bibfnamefont {A.~M.~S.}\ \bibnamefont {Richards}}, \bibinfo {author}
  {\bibfnamefont {P.}~\bibnamefont {Kervella}}, \ and\ \bibinfo {author}
  {\bibfnamefont {L.}~\bibnamefont {Decin}},\ }\bibfield  {title} {\enquote
  {\bibinfo {title} {{An Updated 2017 Astrometric Solution for Betelgeuse}},}\
  }\href {\doibase 10.3847/1538-3881/aa6ff9} {\bibfield  {journal} {\bibinfo
  {journal} {Astron. J.}\ }\textbf {\bibinfo {volume} {154}},\ \bibinfo {pages}
  {11} (\bibinfo {year} {2017})}\BibitemShut {NoStop}%
\bibitem [{\citenamefont {B\"ahre}\ \emph {et~al.}(2013)\citenamefont {B\"ahre}
  \emph {et~al.}}]{alps2:2013ywa}%
  \BibitemOpen
  \bibfield  {author} {\bibinfo {author} {\bibfnamefont {Robin}\ \bibnamefont
  {B\"ahre}} \emph {et~al.},\ }\bibfield  {title} {\enquote {\bibinfo {title}
  {{Any light particle search II -- Technical Design Report}},}\ }\href
  {\doibase 10.1088/1748-0221/8/09/T09001} {\bibfield  {journal} {\bibinfo
  {journal} {JINST}\ }\textbf {\bibinfo {volume} {8}},\ \bibinfo {pages}
  {T09001} (\bibinfo {year} {2013})},\ \Eprint {http://arxiv.org/abs/1302.5647}
  {arXiv:1302.5647 [physics.ins-det]} \BibitemShut {NoStop}%
\bibitem [{\citenamefont {Armengaud}\ \emph {et~al.}(2019)\citenamefont
  {Armengaud} \emph {et~al.}}]{Armengaud:2019uso}%
  \BibitemOpen
  \bibfield  {author} {\bibinfo {author} {\bibfnamefont {E.}~\bibnamefont
  {Armengaud}} \emph {et~al.} (\bibinfo {collaboration} {IAXO}),\ }\bibfield
  {title} {\enquote {\bibinfo {title} {{Physics potential of the International
  Axion Observatory (IAXO)}},}\ }\href {\doibase 10.1088/1475-7516/2019/06/047}
  {\bibfield  {journal} {\bibinfo  {journal} {JCAP}\ }\textbf {\bibinfo
  {volume} {06}},\ \bibinfo {pages} {047} (\bibinfo {year} {2019})},\ \Eprint
  {http://arxiv.org/abs/1904.09155} {arXiv:1904.09155 [hep-ph]} \BibitemShut
  {NoStop}%
\bibitem [{\citenamefont {Di~Vecchia}\ \emph {et~al.}(2019)\citenamefont
  {Di~Vecchia}, \citenamefont {Giannotti}, \citenamefont {Lattanzi},\ and\
  \citenamefont {Lindner}}]{DiVecchia:2019ejf}%
  \BibitemOpen
  \bibfield  {author} {\bibinfo {author} {\bibfnamefont {Paolo}\ \bibnamefont
  {Di~Vecchia}}, \bibinfo {author} {\bibfnamefont {Maurizio}\ \bibnamefont
  {Giannotti}}, \bibinfo {author} {\bibfnamefont {Massimiliano}\ \bibnamefont
  {Lattanzi}}, \ and\ \bibinfo {author} {\bibfnamefont {Axel}\ \bibnamefont
  {Lindner}},\ }\bibfield  {title} {\enquote {\bibinfo {title} {{Round Table on
  Axions and Axion-like Particles}},}\ }\href {\doibase 10.22323/1.336.0034}
  {\bibfield  {journal} {\bibinfo  {journal} {PoS}\ }\textbf {\bibinfo {volume}
  {Confinement2018}},\ \bibinfo {pages} {034} (\bibinfo {year} {2019})},\
  \Eprint {http://arxiv.org/abs/1902.06567} {arXiv:1902.06567 [hep-ph]}
  \BibitemShut {NoStop}%
\bibitem [{\citenamefont {{Straniero}}\ \emph {et~al.}(2019)\citenamefont
  {{Straniero}}, \citenamefont {{Dominguez}}, \citenamefont {{Piersanti}},
  \citenamefont {{Giannotti}},\ and\ \citenamefont
  {{Mirizzi}}}]{Straniero:2019}%
  \BibitemOpen
  \bibfield  {author} {\bibinfo {author} {\bibfnamefont {Oscar}\ \bibnamefont
  {{Straniero}}}, \bibinfo {author} {\bibfnamefont {Inma}\ \bibnamefont
  {{Dominguez}}}, \bibinfo {author} {\bibfnamefont {Luciano}\ \bibnamefont
  {{Piersanti}}}, \bibinfo {author} {\bibfnamefont {Maurizio}\ \bibnamefont
  {{Giannotti}}}, \ and\ \bibinfo {author} {\bibfnamefont {Alessandro}\
  \bibnamefont {{Mirizzi}}},\ }\bibfield  {title} {\enquote {\bibinfo {title}
  {{The Initial Mass-Final Luminosity Relation of Type II Supernova
  Progenitors: Hints of New Physics?}}}\ }\href {\doibase
  10.3847/1538-4357/ab3222} {\bibfield  {journal} {\bibinfo  {journal} {\apj}\
  }\textbf {\bibinfo {volume} {881}},\ \bibinfo {eid} {158} (\bibinfo {year}
  {2019})},\ \Eprint {http://arxiv.org/abs/1907.06367} {arXiv:1907.06367
  [astro-ph.SR]} \BibitemShut {NoStop}%
\bibitem [{\citenamefont {Han}(2017)}]{Han2017}%
  \BibitemOpen
  \bibfield  {author} {\bibinfo {author} {\bibfnamefont {J.~L.}\ \bibnamefont
  {Han}},\ }\bibfield  {title} {\enquote {\bibinfo {title} {Observing
  interstellar and intergalactic magnetic fields},}\ }\href {\doibase
  10.1146/annurev-astro-091916-055221} {\bibfield  {journal} {\bibinfo
  {journal} {Annual Review of Astronomy and Astrophysics}\ }\textbf {\bibinfo
  {volume} {55}},\ \bibinfo {pages} {111--157} (\bibinfo {year}
  {2017})}\BibitemShut {NoStop}%
\bibitem [{\citenamefont {{Jansson}}\ and\ \citenamefont
  {{Farrar}}(2012)}]{Jansson2012}%
  \BibitemOpen
  \bibfield  {author} {\bibinfo {author} {\bibfnamefont {Ronnie}\ \bibnamefont
  {{Jansson}}}\ and\ \bibinfo {author} {\bibfnamefont {Glennys~R.}\
  \bibnamefont {{Farrar}}},\ }\bibfield  {title} {\enquote {\bibinfo {title}
  {{A New Model of the Galactic Magnetic Field}},}\ }\href {\doibase
  10.1088/0004-637X/757/1/14} {\bibfield  {journal} {\bibinfo  {journal}
  {\apj}\ }\textbf {\bibinfo {volume} {757}},\ \bibinfo {eid} {14} (\bibinfo
  {year} {2012})},\ \Eprint {http://arxiv.org/abs/1204.3662} {arXiv:1204.3662
  [astro-ph.GA]} \BibitemShut {NoStop}%
\bibitem [{\citenamefont {Xu}\ and\ \citenamefont {Han}(2019)}]{XuHan2019}%
  \BibitemOpen
  \bibfield  {author} {\bibinfo {author} {\bibfnamefont {J.}~\bibnamefont
  {Xu}}\ and\ \bibinfo {author} {\bibfnamefont {J.~L.}\ \bibnamefont {Han}},\
  }\bibfield  {title} {\enquote {\bibinfo {title} {{Magnetic fields in the
  solar vicinity and in the Galactic halo}},}\ }\href {\doibase
  10.1093/mnras/stz1060} {\bibfield  {journal} {\bibinfo  {journal} {Monthly
  Notices of the Royal Astronomical Society}\ }\textbf {\bibinfo {volume}
  {486}},\ \bibinfo {pages} {4275--4289} (\bibinfo {year} {2019})}\BibitemShut
  {NoStop}%
\bibitem [{\citenamefont {Beck}\ \emph {et~al.}(2016)\citenamefont {Beck},
  \citenamefont {Beck}, \citenamefont {Beck}, \citenamefont {Dolag},
  \citenamefont {Strong},\ and\ \citenamefont {Nielaba}}]{Beck2016}%
  \BibitemOpen
  \bibfield  {author} {\bibinfo {author} {\bibfnamefont {Marcus~C.}\
  \bibnamefont {Beck}}, \bibinfo {author} {\bibfnamefont {Alexander~M.}\
  \bibnamefont {Beck}}, \bibinfo {author} {\bibfnamefont {Rainer}\ \bibnamefont
  {Beck}}, \bibinfo {author} {\bibfnamefont {Klaus}\ \bibnamefont {Dolag}},
  \bibinfo {author} {\bibfnamefont {Andrew~W.}\ \bibnamefont {Strong}}, \ and\
  \bibinfo {author} {\bibfnamefont {Peter}\ \bibnamefont {Nielaba}},\
  }\bibfield  {title} {\enquote {\bibinfo {title} {{New constraints on
  modelling the random magnetic field of the MW}},}\ }\href {\doibase
  10.1088/1475-7516/2016/05/056} {\bibfield  {journal} {\bibinfo  {journal}
  {JCAP}\ }\textbf {\bibinfo {volume} {05}},\ \bibinfo {pages} {056} (\bibinfo
  {year} {2016})},\ \Eprint {http://arxiv.org/abs/1409.5120} {arXiv:1409.5120
  [astro-ph.GA]} \BibitemShut {NoStop}%
\bibitem [{\citenamefont {{Pelgrims}}\ \emph {et~al.}(2020)\citenamefont
  {{Pelgrims}}, \citenamefont {{Ferri{\`e}re}}, \citenamefont {{Boulanger}},
  \citenamefont {{Lallement}},\ and\ \citenamefont {{Montier}}}]{Pelgrims2020}%
  \BibitemOpen
  \bibfield  {author} {\bibinfo {author} {\bibfnamefont {V.}~\bibnamefont
  {{Pelgrims}}}, \bibinfo {author} {\bibfnamefont {K.}~\bibnamefont
  {{Ferri{\`e}re}}}, \bibinfo {author} {\bibfnamefont {F.}~\bibnamefont
  {{Boulanger}}}, \bibinfo {author} {\bibfnamefont {R.}~\bibnamefont
  {{Lallement}}}, \ and\ \bibinfo {author} {\bibfnamefont {L.}~\bibnamefont
  {{Montier}}},\ }\bibfield  {title} {\enquote {\bibinfo {title} {{Modeling the
  magnetized Local Bubble from dust data}},}\ }\href {\doibase
  10.1051/0004-6361/201937157} {\bibfield  {journal} {\bibinfo  {journal}
  {Astronomy and Astrophysics}\ }\textbf {\bibinfo {volume} {636}},\ \bibinfo
  {eid} {A17} (\bibinfo {year} {2020})},\ \Eprint
  {http://arxiv.org/abs/1911.09691} {arXiv:1911.09691 [astro-ph.GA]}
  \BibitemShut {NoStop}%
\bibitem [{\citenamefont {Bassan}\ \emph {et~al.}(2010)\citenamefont {Bassan},
  \citenamefont {Mirizzi},\ and\ \citenamefont {Roncadelli}}]{Bassan:2010ya}%
  \BibitemOpen
  \bibfield  {author} {\bibinfo {author} {\bibfnamefont {Nicola}\ \bibnamefont
  {Bassan}}, \bibinfo {author} {\bibfnamefont {Alessandro}\ \bibnamefont
  {Mirizzi}}, \ and\ \bibinfo {author} {\bibfnamefont {Marco}\ \bibnamefont
  {Roncadelli}},\ }\bibfield  {title} {\enquote {\bibinfo {title} {{Axion-Like
  Particle Effects on the Polarization of Cosmic High-Energy Gamma Sources}},}\
  }\href {\doibase 10.1088/1475-7516/2010/05/010} {\bibfield  {journal}
  {\bibinfo  {journal} {J. Cosmol. Astropart. Phys.}\ }\textbf {\bibinfo
  {volume} {05}},\ \bibinfo {pages} {010} (\bibinfo {year} {2010})},\ \Eprint
  {http://arxiv.org/abs/1001.5267} {arXiv:1001.5267 [astro-ph.HE]} \BibitemShut
  {NoStop}%
\bibitem [{\citenamefont {Cordes}\ and\ \citenamefont
  {Lazio}(2002)}]{Cordes:2002wz}%
  \BibitemOpen
  \bibfield  {author} {\bibinfo {author} {\bibfnamefont {James~M.}\
  \bibnamefont {Cordes}}\ and\ \bibinfo {author} {\bibfnamefont {T.J.W.}\
  \bibnamefont {Lazio}},\ }\bibfield  {title} {\enquote {\bibinfo {title}
  {{NE2001. I. A New model for the galactic distribution of free electrons and
  its fluctuations}},}\ }\href@noop {} {\  (\bibinfo {year} {2002})},\ \Eprint
  {http://arxiv.org/abs/astro-ph/0207156} {arXiv:astro-ph/0207156} \BibitemShut
  {NoStop}%
\bibitem [{\citenamefont {Harvey-Smith}\ \emph {et~al.}(2011)\citenamefont
  {Harvey-Smith}, \citenamefont {Madsen},\ and\ \citenamefont
  {Gaensler}}]{HarveySmith:2011fe}%
  \BibitemOpen
  \bibfield  {author} {\bibinfo {author} {\bibfnamefont {Lisa}\ \bibnamefont
  {Harvey-Smith}}, \bibinfo {author} {\bibfnamefont {Gregory~J.}\ \bibnamefont
  {Madsen}}, \ and\ \bibinfo {author} {\bibfnamefont {Bryan~M.}\ \bibnamefont
  {Gaensler}},\ }\bibfield  {title} {\enquote {\bibinfo {title} {{Magnetic
  Fields in Large Diameter H\textsc{ii} Regions Revealed by the Faraday
  Rotation of Compact Extragalactic Radio Sources}},}\ }\href {\doibase
  10.1088/0004-637X/736/2/83} {\bibfield  {journal} {\bibinfo  {journal}
  {Astrophys. J.}\ }\textbf {\bibinfo {volume} {736}},\ \bibinfo {pages} {83}
  (\bibinfo {year} {2011})},\ \Eprint {http://arxiv.org/abs/1106.0931}
  {arXiv:1106.0931 [astro-ph.GA]} \BibitemShut {NoStop}%
\bibitem [{\citenamefont {Harrison}\ \emph {et~al.}(2013)\citenamefont
  {Harrison} \emph {et~al.}}]{Harrison:2013md}%
  \BibitemOpen
  \bibfield  {author} {\bibinfo {author} {\bibfnamefont {Fiona~A.}\
  \bibnamefont {Harrison}} \emph {et~al.},\ }\bibfield  {title} {\enquote
  {\bibinfo {title} {{The Nuclear Spectroscopic Telescope Array (NuSTAR)
  High-Energy X-Ray Mission}},}\ }\href {\doibase 10.1088/0004-637X/770/2/103}
  {\bibfield  {journal} {\bibinfo  {journal} {Astrophys. J.}\ }\textbf
  {\bibinfo {volume} {770}},\ \bibinfo {pages} {103} (\bibinfo {year}
  {2013})},\ \Eprint {http://arxiv.org/abs/1301.7307} {arXiv:1301.7307
  [astro-ph.IM]} \BibitemShut {NoStop}%
\bibitem [{\citenamefont {Madsen}\ \emph {et~al.}(2015)\citenamefont {Madsen}
  \emph {et~al.}}]{Madsen:2015jea}%
  \BibitemOpen
  \bibfield  {author} {\bibinfo {author} {\bibfnamefont {Kristin~K.}\
  \bibnamefont {Madsen}} \emph {et~al.},\ }\bibfield  {title} {\enquote
  {\bibinfo {title} {{Calibration of the NuSTAR High Energy Focusing X-ray
  Telescope}},}\ }\href {\doibase 10.1088/0067-0049/220/1/8} {\bibfield
  {journal} {\bibinfo  {journal} {Astrophys. J. Suppl.}\ }\textbf {\bibinfo
  {volume} {220}},\ \bibinfo {pages} {8} (\bibinfo {year} {2015})},\ \Eprint
  {http://arxiv.org/abs/1504.01672} {arXiv:1504.01672 [astro-ph.IM]}
  \BibitemShut {NoStop}%
\bibitem [{\citenamefont {Perez}\ \emph {et~al.}(2019)\citenamefont {Perez},
  \citenamefont {Krivonos},\ and\ \citenamefont {Wik}}]{Perez_2019}%
  \BibitemOpen
  \bibfield  {author} {\bibinfo {author} {\bibfnamefont {Kerstin}\ \bibnamefont
  {Perez}}, \bibinfo {author} {\bibfnamefont {Roman}\ \bibnamefont {Krivonos}},
  \ and\ \bibinfo {author} {\bibfnamefont {Daniel~R.}\ \bibnamefont {Wik}},\
  }\bibfield  {title} {\enquote {\bibinfo {title} {{The Galactic Bulge Diffuse
  Emission in Broadband X-Rays with {NuSTAR}}},}\ }\href {\doibase
  10.3847/1538-4357/ab4590} {\bibfield  {journal} {\bibinfo  {journal}
  {Astrophys. J.}\ }\textbf {\bibinfo {volume} {884}},\ \bibinfo {pages} {153}
  (\bibinfo {year} {2019})},\ \Eprint {http://arxiv.org/abs/1909.05916}
  {arXiv:1909.05916 [astro-ph.HE]} \BibitemShut {NoStop}%
\bibitem [{\citenamefont {Roach}\ \emph {et~al.}(2020)\citenamefont {Roach},
  \citenamefont {Ng}, \citenamefont {Perez}, \citenamefont {Beacom},
  \citenamefont {Horiuchi}, \citenamefont {Krivonos},\ and\ \citenamefont
  {Wik}}]{Roach:2019ctw}%
  \BibitemOpen
  \bibfield  {author} {\bibinfo {author} {\bibfnamefont {Brandon~M.}\
  \bibnamefont {Roach}}, \bibinfo {author} {\bibfnamefont {Kenny C.~Y.}\
  \bibnamefont {Ng}}, \bibinfo {author} {\bibfnamefont {Kerstin}\ \bibnamefont
  {Perez}}, \bibinfo {author} {\bibfnamefont {John~F.}\ \bibnamefont {Beacom}},
  \bibinfo {author} {\bibfnamefont {Shunsaku}\ \bibnamefont {Horiuchi}},
  \bibinfo {author} {\bibfnamefont {Roman}\ \bibnamefont {Krivonos}}, \ and\
  \bibinfo {author} {\bibfnamefont {Daniel~R.}\ \bibnamefont {Wik}},\
  }\bibfield  {title} {\enquote {\bibinfo {title} {{NuSTAR Tests of
  Sterile-Neutrino Dark Matter: New Galactic Bulge Observations and Combined
  Impact}},}\ }\href {\doibase 10.1103/PhysRevD.101.103011} {\bibfield
  {journal} {\bibinfo  {journal} {Phys. Rev. D}\ }\textbf {\bibinfo {volume}
  {101}},\ \bibinfo {pages} {103011} (\bibinfo {year} {2020})},\ \Eprint
  {http://arxiv.org/abs/1908.09037} {arXiv:1908.09037 [astro-ph.HE]}
  \BibitemShut {NoStop}%
\bibitem [{\citenamefont {van Leeuwen}(2007)}]{Hipparcos}%
  \BibitemOpen
  \bibfield  {author} {\bibinfo {author} {\bibfnamefont {F.}~\bibnamefont {van
  Leeuwen}},\ }\bibfield  {title} {\enquote {\bibinfo {title} {{Validation of
  the New Hipparcos Reduction}},}\ }\href {\doibase 10.1051/0004-6361:20078357}
  {\bibfield  {journal} {\bibinfo  {journal} {Astron. Astrophys.}\ }\textbf
  {\bibinfo {volume} {474}},\ \bibinfo {pages} {653--664} (\bibinfo {year}
  {2007})}\BibitemShut {NoStop}%
\bibitem [{\citenamefont {Evans}\ \emph {et~al.}(2010)\citenamefont {Evans}
  \emph {et~al.}}]{Evans_2010}%
  \BibitemOpen
  \bibfield  {author} {\bibinfo {author} {\bibfnamefont {I.~N.}\ \bibnamefont
  {Evans}} \emph {et~al.},\ }\bibfield  {title} {\enquote {\bibinfo {title}
  {{The Chandra Source Catalog}},}\ }\href {\doibase
  10.1088/0067-0049/189/1/37} {\bibfield  {journal} {\bibinfo  {journal}
  {Astrophys. J. Suppl. Ser.}\ }\textbf {\bibinfo {volume} {189}},\ \bibinfo
  {pages} {37--82} (\bibinfo {year} {2010})},\ \Eprint
  {http://arxiv.org/abs/1005.4665} {arXiv:1005.4665 [astro-ph.HE]} \BibitemShut
  {NoStop}%
\bibitem [{\citenamefont {Wik}\ \emph {et~al.}(2014)\citenamefont {Wik} \emph
  {et~al.}}]{Wik:2014}%
  \BibitemOpen
  \bibfield  {author} {\bibinfo {author} {\bibfnamefont {D.~R.}\ \bibnamefont
  {Wik}} \emph {et~al.},\ }\bibfield  {title} {\enquote {\bibinfo {title}
  {{{NuSTAR Observations of the Bullet Cluster: Constraints on Inverse Compton
  Emission}}},}\ }\href {\doibase 10.1088/0004-637X/792/1/48} {\bibfield
  {journal} {\bibinfo  {journal} {Astrophys. J.}\ }\textbf {\bibinfo {volume}
  {792}},\ \bibinfo {pages} {48} (\bibinfo {year} {2014})},\ \Eprint
  {http://arxiv.org/abs/1403.2722} {arXiv:1403.2722 [astro-ph.HE]} \BibitemShut
  {NoStop}%
\bibitem [{\citenamefont {Brun}\ and\ \citenamefont
  {Rademakers}(1997)}]{Brun:1997pa}%
  \BibitemOpen
  \bibfield  {author} {\bibinfo {author} {\bibfnamefont {R.}~\bibnamefont
  {Brun}}\ and\ \bibinfo {author} {\bibfnamefont {F.}~\bibnamefont
  {Rademakers}},\ }\bibfield  {title} {\enquote {\bibinfo {title} {{ROOT: An
  object oriented data analysis framework}},}\ }\href {\doibase
  10.1016/S0168-9002(97)00048-X} {\bibfield  {journal} {\bibinfo  {journal}
  {Nucl. Instrum. Meth. A}\ }\textbf {\bibinfo {volume} {389}},\ \bibinfo
  {pages} {81--86} (\bibinfo {year} {1997})}\BibitemShut {NoStop}%
\bibitem [{\citenamefont {Junk}(1999)}]{Junk:1999kv}%
  \BibitemOpen
  \bibfield  {author} {\bibinfo {author} {\bibfnamefont {Thomas}\ \bibnamefont
  {Junk}},\ }\bibfield  {title} {\enquote {\bibinfo {title} {{Confidence Level
  Computation for Combining Searches with Small Statistics}},}\ }\href
  {\doibase 10.1016/S0168-9002(99)00498-2} {\bibfield  {journal} {\bibinfo
  {journal} {Nucl. Instrum. Meth. A}\ }\textbf {\bibinfo {volume} {434}},\
  \bibinfo {pages} {435--443} (\bibinfo {year} {1999})},\ \Eprint
  {http://arxiv.org/abs/hep-ex/9902006} {arXiv:hep-ex/9902006} \BibitemShut
  {NoStop}%
\bibitem [{\citenamefont {Cowan}\ \emph {et~al.}(2011)\citenamefont {Cowan},
  \citenamefont {Cranmer}, \citenamefont {Gross},\ and\ \citenamefont
  {Vitells}}]{Cowan:2010js}%
  \BibitemOpen
  \bibfield  {author} {\bibinfo {author} {\bibfnamefont {Glen}\ \bibnamefont
  {Cowan}}, \bibinfo {author} {\bibfnamefont {Kyle}\ \bibnamefont {Cranmer}},
  \bibinfo {author} {\bibfnamefont {Eilam}\ \bibnamefont {Gross}}, \ and\
  \bibinfo {author} {\bibfnamefont {Ofer}\ \bibnamefont {Vitells}},\ }\bibfield
   {title} {\enquote {\bibinfo {title} {{Asymptotic Formulae for
  Likelihood-Based Tests of New Physics}},}\ }\href {\doibase
  10.1140/epjc/s10052-011-1554-0} {\bibfield  {journal} {\bibinfo  {journal}
  {Eur. Phys. J. C}\ }\textbf {\bibinfo {volume} {71}},\ \bibinfo {pages}
  {1554} (\bibinfo {year} {2011})},\ \bibinfo {note} {[Erratum: Eur.Phys.J.C
  73, 2501 (2013)]},\ \Eprint {http://arxiv.org/abs/1007.1727} {arXiv:1007.1727
  [physics.data-an]} \BibitemShut {NoStop}%
\bibitem [{\citenamefont {Feldman}\ and\ \citenamefont
  {Cousins}(1998)}]{Feldman:1997qc}%
  \BibitemOpen
  \bibfield  {author} {\bibinfo {author} {\bibfnamefont {Gary~J.}\ \bibnamefont
  {Feldman}}\ and\ \bibinfo {author} {\bibfnamefont {Robert~D.}\ \bibnamefont
  {Cousins}},\ }\bibfield  {title} {\enquote {\bibinfo {title} {{A Unified
  Approach to the Classical Statistical Analysis of Small Signals}},}\ }\href
  {\doibase 10.1103/PhysRevD.57.3873} {\bibfield  {journal} {\bibinfo
  {journal} {Phys. Rev. D}\ }\textbf {\bibinfo {volume} {57}},\ \bibinfo
  {pages} {3873--3889} (\bibinfo {year} {1998})},\ \Eprint
  {http://arxiv.org/abs/physics/9711021} {arXiv:physics/9711021} \BibitemShut
  {NoStop}%
\bibitem [{\citenamefont {Ouellet}\ \emph
  {et~al.}(2019{\natexlab{a}})\citenamefont {Ouellet} \emph
  {et~al.}}]{Ouellet:2018beu}%
  \BibitemOpen
  \bibfield  {author} {\bibinfo {author} {\bibfnamefont {Jonathan~L.}\
  \bibnamefont {Ouellet}} \emph {et~al.},\ }\bibfield  {title} {\enquote
  {\bibinfo {title} {{First Results from ABRACADABRA-10 cm: A Search for
  Sub-$\mu$eV Axion Dark Matter}},}\ }\href {\doibase
  10.1103/PhysRevLett.122.121802} {\bibfield  {journal} {\bibinfo  {journal}
  {Phys. Rev. Lett.}\ }\textbf {\bibinfo {volume} {122}},\ \bibinfo {pages}
  {121802} (\bibinfo {year} {2019}{\natexlab{a}})},\ \Eprint
  {http://arxiv.org/abs/1810.12257} {arXiv:1810.12257 [hep-ex]} \BibitemShut
  {NoStop}%
\bibitem [{\citenamefont {Braine}\ \emph {et~al.}(2020)\citenamefont {Braine}
  \emph {et~al.}}]{admx:2019fqb}%
  \BibitemOpen
  \bibfield  {author} {\bibinfo {author} {\bibfnamefont {T.}~\bibnamefont
  {Braine}} \emph {et~al.} (\bibinfo {collaboration} {ADMX}),\ }\bibfield
  {title} {\enquote {\bibinfo {title} {{Extended Search for the Invisible Axion
  with the Axion Dark Matter Experiment}},}\ }\href {\doibase
  10.1103/PhysRevLett.124.101303} {\bibfield  {journal} {\bibinfo  {journal}
  {Phys. Rev. Lett.}\ }\textbf {\bibinfo {volume} {124}},\ \bibinfo {pages}
  {101303} (\bibinfo {year} {2020})},\ \Eprint
  {http://arxiv.org/abs/1910.08638} {arXiv:1910.08638 [hep-ex]} \BibitemShut
  {NoStop}%
\bibitem [{\citenamefont {Di~Luzio}\ \emph {et~al.}(2017)\citenamefont
  {Di~Luzio}, \citenamefont {Mescia},\ and\ \citenamefont
  {Nardi}}]{DiLuzio:2016sbl}%
  \BibitemOpen
  \bibfield  {author} {\bibinfo {author} {\bibfnamefont {Luca}\ \bibnamefont
  {Di~Luzio}}, \bibinfo {author} {\bibfnamefont {Federico}\ \bibnamefont
  {Mescia}}, \ and\ \bibinfo {author} {\bibfnamefont {Enrico}\ \bibnamefont
  {Nardi}},\ }\bibfield  {title} {\enquote {\bibinfo {title} {{Redefining the
  Axion Window}},}\ }\href {\doibase 10.1103/PhysRevLett.118.031801} {\bibfield
   {journal} {\bibinfo  {journal} {Phys. Rev. Lett.}\ }\textbf {\bibinfo
  {volume} {118}},\ \bibinfo {pages} {031801} (\bibinfo {year} {2017})},\
  \Eprint {http://arxiv.org/abs/1610.07593} {arXiv:1610.07593 [hep-ph]}
  \BibitemShut {NoStop}%
\bibitem [{\citenamefont {Giannotti}\ \emph {et~al.}(2016)\citenamefont
  {Giannotti}, \citenamefont {Irastorza}, \citenamefont {Redondo},\ and\
  \citenamefont {Ringwald}}]{Giannotti:2015kwo}%
  \BibitemOpen
  \bibfield  {author} {\bibinfo {author} {\bibfnamefont {Maurizio}\
  \bibnamefont {Giannotti}}, \bibinfo {author} {\bibfnamefont {Igor}\
  \bibnamefont {Irastorza}}, \bibinfo {author} {\bibfnamefont {Javier}\
  \bibnamefont {Redondo}}, \ and\ \bibinfo {author} {\bibfnamefont {Andreas}\
  \bibnamefont {Ringwald}},\ }\bibfield  {title} {\enquote {\bibinfo {title}
  {{Cool WISPs for stellar cooling excesses}},}\ }\href {\doibase
  10.1088/1475-7516/2016/05/057} {\bibfield  {journal} {\bibinfo  {journal}
  {JCAP}\ }\textbf {\bibinfo {volume} {05}},\ \bibinfo {pages} {057} (\bibinfo
  {year} {2016})},\ \Eprint {http://arxiv.org/abs/1512.08108} {arXiv:1512.08108
  [astro-ph.HE]} \BibitemShut {NoStop}%
\bibitem [{\citenamefont {Giannotti}\ \emph {et~al.}(2017)\citenamefont
  {Giannotti}, \citenamefont {Irastorza}, \citenamefont {Redondo},
  \citenamefont {Ringwald},\ and\ \citenamefont {Saikawa}}]{Giannotti:2017hny}%
  \BibitemOpen
  \bibfield  {author} {\bibinfo {author} {\bibfnamefont {Maurizio}\
  \bibnamefont {Giannotti}}, \bibinfo {author} {\bibfnamefont {Igor~G.}\
  \bibnamefont {Irastorza}}, \bibinfo {author} {\bibfnamefont {Javier}\
  \bibnamefont {Redondo}}, \bibinfo {author} {\bibfnamefont {Andreas}\
  \bibnamefont {Ringwald}}, \ and\ \bibinfo {author} {\bibfnamefont {Ken'ichi}\
  \bibnamefont {Saikawa}},\ }\bibfield  {title} {\enquote {\bibinfo {title}
  {{Stellar Recipes for Axion Hunters}},}\ }\href {\doibase
  10.1088/1475-7516/2017/10/010} {\bibfield  {journal} {\bibinfo  {journal}
  {JCAP}\ }\textbf {\bibinfo {volume} {10}},\ \bibinfo {pages} {010} (\bibinfo
  {year} {2017})},\ \Eprint {http://arxiv.org/abs/1708.02111} {arXiv:1708.02111
  [hep-ph]} \BibitemShut {NoStop}%
\bibitem [{\citenamefont {{Salvati}}(2010)}]{Salvati2010}%
  \BibitemOpen
  \bibfield  {author} {\bibinfo {author} {\bibfnamefont {M.}~\bibnamefont
  {{Salvati}}},\ }\bibfield  {title} {\enquote {\bibinfo {title} {{The local
  Galactic magnetic field in the direction of Geminga}},}\ }\href {\doibase
  10.1051/0004-6361/200913406} {\bibfield  {journal} {\bibinfo  {journal}
  {Astronomy and Astrophysics}\ }\textbf {\bibinfo {volume} {513}},\ \bibinfo
  {eid} {A28} (\bibinfo {year} {2010})},\ \Eprint
  {http://arxiv.org/abs/1001.4947} {arXiv:1001.4947 [astro-ph.HE]} \BibitemShut
  {NoStop}%
\bibitem [{\citenamefont {Frisch}\ \emph {et~al.}(2012)\citenamefont {Frisch},
  \citenamefont {Andersson}, \citenamefont {Berdyugin}, \citenamefont
  {Piirola}, \citenamefont {DeMajistre}, \citenamefont {Funsten}, \citenamefont
  {Magalhaes}, \citenamefont {Seriacopi}, \citenamefont {McComas},
  \citenamefont {Schwadron}, \citenamefont {Slavin},\ and\ \citenamefont
  {Wiktorowicz}}]{Frisch2012}%
  \BibitemOpen
  \bibfield  {author} {\bibinfo {author} {\bibfnamefont {P.~C.}\ \bibnamefont
  {Frisch}}, \bibinfo {author} {\bibfnamefont {B-G}\ \bibnamefont {Andersson}},
  \bibinfo {author} {\bibfnamefont {A.}~\bibnamefont {Berdyugin}}, \bibinfo
  {author} {\bibfnamefont {V.}~\bibnamefont {Piirola}}, \bibinfo {author}
  {\bibfnamefont {R.}~\bibnamefont {DeMajistre}}, \bibinfo {author}
  {\bibfnamefont {H.~O.}\ \bibnamefont {Funsten}}, \bibinfo {author}
  {\bibfnamefont {A.~M.}\ \bibnamefont {Magalhaes}}, \bibinfo {author}
  {\bibfnamefont {D.~B.}\ \bibnamefont {Seriacopi}}, \bibinfo {author}
  {\bibfnamefont {D.~J.}\ \bibnamefont {McComas}}, \bibinfo {author}
  {\bibfnamefont {N.~A.}\ \bibnamefont {Schwadron}}, \bibinfo {author}
  {\bibfnamefont {J.~D.}\ \bibnamefont {Slavin}}, \ and\ \bibinfo {author}
  {\bibfnamefont {S.~J.}\ \bibnamefont {Wiktorowicz}},\ }\bibfield  {title}
  {\enquote {\bibinfo {title} {{The} {Interstellar} {Magnetic} {Field} {Close}
  {to} {the} {SUN}. {II}.}}\ }\href {\doibase 10.1088/0004-637x/760/2/106}
  {\bibfield  {journal} {\bibinfo  {journal} {The Astrophysical Journal}\
  }\textbf {\bibinfo {volume} {760}},\ \bibinfo {pages} {106} (\bibinfo {year}
  {2012})}\BibitemShut {NoStop}%
\bibitem [{\citenamefont {Ouellet}\ \emph
  {et~al.}(2019{\natexlab{b}})\citenamefont {Ouellet} \emph
  {et~al.}}]{Ouellet:2019tlz}%
  \BibitemOpen
  \bibfield  {author} {\bibinfo {author} {\bibfnamefont {Jonathan~L.}\
  \bibnamefont {Ouellet}} \emph {et~al.},\ }\bibfield  {title} {\enquote
  {\bibinfo {title} {{Design and Implementation of the ABRACADABRA-10 cm Axion
  Dark Matter Search}},}\ }\href {\doibase 10.1103/PhysRevD.99.052012}
  {\bibfield  {journal} {\bibinfo  {journal} {Phys. Rev. D}\ }\textbf {\bibinfo
  {volume} {99}},\ \bibinfo {pages} {052012} (\bibinfo {year}
  {2019}{\natexlab{b}})},\ \Eprint {http://arxiv.org/abs/1901.10652}
  {arXiv:1901.10652 [physics.ins-det]} \BibitemShut {NoStop}%
\bibitem [{\citenamefont {Silva-Feaver}\ \emph {et~al.}(2017)\citenamefont
  {Silva-Feaver} \emph {et~al.}}]{dm_radio}%
  \BibitemOpen
  \bibfield  {author} {\bibinfo {author} {\bibfnamefont {M.}~\bibnamefont
  {Silva-Feaver}} \emph {et~al.},\ }\bibfield  {title} {\enquote {\bibinfo
  {title} {{Design Overview of DM Radio Pathfinder Experiment}},}\ }\href
  {\doibase 10.1109/TASC.2016.2631425} {\bibfield  {journal} {\bibinfo
  {journal} {{IEEE Transactions on Applied Superconductivity}}\ }\textbf
  {\bibinfo {volume} {27}},\ \bibinfo {pages} {1--4} (\bibinfo {year}
  {2017})}\BibitemShut {NoStop}%
\bibitem [{\citenamefont {Fabian}\ and\ \citenamefont
  {Barcons}(1992)}]{Fabian:1992}%
  \BibitemOpen
  \bibfield  {author} {\bibinfo {author} {\bibfnamefont {A.~C.}\ \bibnamefont
  {Fabian}}\ and\ \bibinfo {author} {\bibfnamefont {X.}~\bibnamefont
  {Barcons}},\ }\bibfield  {title} {\enquote {\bibinfo {title} {{The Origin of
  the X-Ray Background}},}\ }\href {\doibase
  10.1146/annurev.aa.30.090192.002241} {\bibfield  {journal} {\bibinfo
  {journal} {Annu. Rev. Astron. Astrophys.}\ }\textbf {\bibinfo {volume}
  {30}},\ \bibinfo {pages} {429--456} (\bibinfo {year} {1992})}\BibitemShut
  {NoStop}%
\bibitem [{\citenamefont {Revnivtsev}\ \emph {et~al.}(2003)\citenamefont
  {Revnivtsev}, \citenamefont {Gilfanov}, \citenamefont {Sunyaev},
  \citenamefont {Jahoda},\ and\ \citenamefont {Markwardt}}]{Revnivtsev:2003wm}%
  \BibitemOpen
  \bibfield  {author} {\bibinfo {author} {\bibfnamefont {Mikhail}\ \bibnamefont
  {Revnivtsev}}, \bibinfo {author} {\bibfnamefont {M.}~\bibnamefont
  {Gilfanov}}, \bibinfo {author} {\bibfnamefont {R.}~\bibnamefont {Sunyaev}},
  \bibinfo {author} {\bibfnamefont {K.}~\bibnamefont {Jahoda}}, \ and\ \bibinfo
  {author} {\bibfnamefont {C.}~\bibnamefont {Markwardt}},\ }\bibfield  {title}
  {\enquote {\bibinfo {title} {{The Spectrum of the Cosmic X-Ray Background
  Observed by RTXE/PCA}},}\ }\href {\doibase 10.1051/0004-6361:20031386}
  {\bibfield  {journal} {\bibinfo  {journal} {Astron. Astrophys.}\ }\textbf
  {\bibinfo {volume} {411}},\ \bibinfo {pages} {329--334} (\bibinfo {year}
  {2003})},\ \Eprint {http://arxiv.org/abs/astro-ph/0306569}
  {arXiv:astro-ph/0306569} \BibitemShut {NoStop}%
\bibitem [{\citenamefont {Fruscione}\ \emph {et~al.}(2006)\citenamefont
  {Fruscione}, \citenamefont {McDowell}, \citenamefont {Allen} \emph
  {et~al.}}]{Fruscione06}%
  \BibitemOpen
  \bibfield  {author} {\bibinfo {author} {\bibfnamefont {Antonella}\
  \bibnamefont {Fruscione}}, \bibinfo {author} {\bibfnamefont {Jonathan~C.}\
  \bibnamefont {McDowell}}, \bibinfo {author} {\bibfnamefont {Glenn~E.}\
  \bibnamefont {Allen}},  \emph {et~al.},\ }\bibfield  {title} {\enquote
  {\bibinfo {title} {{CIAO: Chandra's data analysis system}},}\ }\href
  {\doibase 10.1117/12.671760} {\bibfield  {journal} {\bibinfo  {journal}
  {Proc. SPIE, 6270, 62701V}\ } (\bibinfo {year} {2006}),\
  10.1117/12.671760}\BibitemShut {NoStop}%
\bibitem [{\citenamefont {Andriamonje}\ \emph {et~al.}(2007)\citenamefont
  {Andriamonje} \emph {et~al.}}]{Andriamonje:2007ew}%
  \BibitemOpen
  \bibfield  {author} {\bibinfo {author} {\bibfnamefont {S.}~\bibnamefont
  {Andriamonje}} \emph {et~al.} (\bibinfo {collaboration} {CAST}),\ }\bibfield
  {title} {\enquote {\bibinfo {title} {{An Improved limit on the axion-photon
  coupling from the CAST experiment}},}\ }\href {\doibase
  10.1088/1475-7516/2007/04/010} {\bibfield  {journal} {\bibinfo  {journal}
  {JCAP}\ }\textbf {\bibinfo {volume} {04}},\ \bibinfo {pages} {010} (\bibinfo
  {year} {2007})},\ \Eprint {http://arxiv.org/abs/hep-ex/0702006}
  {arXiv:hep-ex/0702006} \BibitemShut {NoStop}%
\bibitem [{\citenamefont {{Le Bertre}}\ \emph {et~al.}(2012)\citenamefont {{Le
  Bertre}}, \citenamefont {{Matthews}}, \citenamefont {{G{\'e}rard}},\ and\
  \citenamefont {{Libert}}}]{Bertre:2012bh}%
  \BibitemOpen
  \bibfield  {author} {\bibinfo {author} {\bibfnamefont {T.}~\bibnamefont {{Le
  Bertre}}}, \bibinfo {author} {\bibfnamefont {L.~D.}\ \bibnamefont
  {{Matthews}}}, \bibinfo {author} {\bibfnamefont {E.}~\bibnamefont
  {{G{\'e}rard}}}, \ and\ \bibinfo {author} {\bibfnamefont {Y.}~\bibnamefont
  {{Libert}}},\ }\bibfield  {title} {\enquote {\bibinfo {title} {{Discovery of
  a detached H I gas shell surrounding {\ensuremath{\alpha}} Orionis}},}\
  }\href {\doibase 10.1111/j.1365-2966.2012.20853.x} {\bibfield  {journal}
  {\bibinfo  {journal} {Mon. Not. R. Astron. Soc}\ }\textbf {\bibinfo {volume}
  {422}},\ \bibinfo {pages} {3433--3443} (\bibinfo {year} {2012})},\ \Eprint
  {http://arxiv.org/abs/1203.0255} {arXiv:1203.0255 [astro-ph.SR]} \BibitemShut
  {NoStop}%
\bibitem [{\citenamefont {{Perrin}}\ \emph {et~al.}(2004)\citenamefont
  {{Perrin}}, \citenamefont {{Ridgway}}, \citenamefont {{Coud{\'e} du
  Foresto}}, \citenamefont {{Mennesson}}, \citenamefont {{Traub}},\ and\
  \citenamefont {{Lacasse}}}]{Perrin:2004ce}%
  \BibitemOpen
  \bibfield  {author} {\bibinfo {author} {\bibfnamefont {G.}~\bibnamefont
  {{Perrin}}}, \bibinfo {author} {\bibfnamefont {S.~T.}\ \bibnamefont
  {{Ridgway}}}, \bibinfo {author} {\bibfnamefont {V.}~\bibnamefont {{Coud{\'e}
  du Foresto}}}, \bibinfo {author} {\bibfnamefont {B.}~\bibnamefont
  {{Mennesson}}}, \bibinfo {author} {\bibfnamefont {W.~A.}\ \bibnamefont
  {{Traub}}}, \ and\ \bibinfo {author} {\bibfnamefont {M.~G.}\ \bibnamefont
  {{Lacasse}}},\ }\bibfield  {title} {\enquote {\bibinfo {title}
  {{Interferometric observations of the supergiant stars {\ensuremath{\alpha}}
  Orionis and {\ensuremath{\alpha}} Herculis with FLUOR at IOTA}},}\ }\href
  {\doibase 10.1051/0004-6361:20040052} {\bibfield  {journal} {\bibinfo
  {journal} {Astron. Astrophys}\ }\textbf {\bibinfo {volume} {418}},\ \bibinfo
  {pages} {675--685} (\bibinfo {year} {2004})},\ \Eprint
  {http://arxiv.org/abs/astro-ph/0402099} {arXiv:astro-ph/0402099 [astro-ph]}
  \BibitemShut {NoStop}%
\bibitem [{\citenamefont {Lambert}\ \emph {et~al.}(1984)\citenamefont
  {Lambert}, \citenamefont {Brown}, \citenamefont {Hinkle},\ and\ \citenamefont
  {Johnson}}]{Lambert:1984}%
  \BibitemOpen
  \bibfield  {author} {\bibinfo {author} {\bibfnamefont {D.~L.}\ \bibnamefont
  {Lambert}}, \bibinfo {author} {\bibfnamefont {J.~A.}\ \bibnamefont {Brown}},
  \bibinfo {author} {\bibfnamefont {K.~H.}\ \bibnamefont {Hinkle}}, \ and\
  \bibinfo {author} {\bibnamefont {Johnson}},\ }\bibfield  {title} {\enquote
  {\bibinfo {title} {{Carbon, nitrogen and oxygem abundances in Betelgeuse}},}\
  }\href {\doibase 10.1086/162401} {\bibfield  {journal} {\bibinfo  {journal}
  {Astrophysical Journal}\ }\textbf {\bibinfo {volume} {284}},\ \bibinfo
  {pages} {223--237} (\bibinfo {year} {1984})}\BibitemShut {NoStop}%
\bibitem [{\citenamefont {{Meynet}}\ \emph {et~al.}(2013)\citenamefont
  {{Meynet}}, \citenamefont {{Haemmerl{\'e}}}, \citenamefont {{Ekstr{\"o}m}},
  \citenamefont {{Georgy}}, \citenamefont {{Groh}},\ and\ \citenamefont
  {{Maeder}}}]{Meynet2013}%
  \BibitemOpen
  \bibfield  {author} {\bibinfo {author} {\bibfnamefont {G.}~\bibnamefont
  {{Meynet}}}, \bibinfo {author} {\bibfnamefont {L.}~\bibnamefont
  {{Haemmerl{\'e}}}}, \bibinfo {author} {\bibfnamefont {S.}~\bibnamefont
  {{Ekstr{\"o}m}}}, \bibinfo {author} {\bibfnamefont {C.}~\bibnamefont
  {{Georgy}}}, \bibinfo {author} {\bibfnamefont {J.}~\bibnamefont {{Groh}}}, \
  and\ \bibinfo {author} {\bibfnamefont {A.}~\bibnamefont {{Maeder}}},\
  }\bibfield  {title} {\enquote {\bibinfo {title} {{The past and future
  evolution of a star like Betelgeuse}},}\ }in\ \href {\doibase
  10.1051/eas/1360002} {\emph {\bibinfo {booktitle} {EAS Publications
  Series}}},\ \bibinfo {series} {EAS Publications Series}, Vol.~\bibinfo
  {volume} {60},\ \bibinfo {editor} {edited by\ \bibinfo {editor}
  {\bibfnamefont {P.}~\bibnamefont {{Kervella}}}, \bibinfo {editor}
  {\bibfnamefont {T.}~\bibnamefont {{Le Bertre}}}, \ and\ \bibinfo {editor}
  {\bibfnamefont {G.}~\bibnamefont {{Perrin}}}}\ (\bibinfo {year} {2013})\ pp.\
  \bibinfo {pages} {17--28},\ \Eprint {http://arxiv.org/abs/1303.1339}
  {arXiv:1303.1339 [astro-ph.SR]} \BibitemShut {NoStop}%
\bibitem [{\citenamefont {{Dolan}}\ \emph {et~al.}(2016)\citenamefont
  {{Dolan}}, \citenamefont {{Mathews}}, \citenamefont {{Lam}}, \citenamefont
  {{Quynh Lan}}, \citenamefont {{Herczeg}},\ and\ \citenamefont
  {{Dearborn}}}]{Dolan2016}%
  \BibitemOpen
  \bibfield  {author} {\bibinfo {author} {\bibfnamefont {Michelle~M.}\
  \bibnamefont {{Dolan}}}, \bibinfo {author} {\bibfnamefont {Grant~J.}\
  \bibnamefont {{Mathews}}}, \bibinfo {author} {\bibfnamefont {Doan~Duc}\
  \bibnamefont {{Lam}}}, \bibinfo {author} {\bibfnamefont {Nguyen}\
  \bibnamefont {{Quynh Lan}}}, \bibinfo {author} {\bibfnamefont {Gregory~J.}\
  \bibnamefont {{Herczeg}}}, \ and\ \bibinfo {author} {\bibfnamefont {David
  S.~P.}\ \bibnamefont {{Dearborn}}},\ }\bibfield  {title} {\enquote {\bibinfo
  {title} {{Evolutionary Tracks for Betelgeuse}},}\ }\href {\doibase
  10.3847/0004-637X/819/1/7} {\bibfield  {journal} {\bibinfo  {journal} {\apj}\
  }\textbf {\bibinfo {volume} {819}},\ \bibinfo {eid} {7} (\bibinfo {year}
  {2016})},\ \Eprint {http://arxiv.org/abs/1406.3143} {arXiv:1406.3143
  [astro-ph.SR]} \BibitemShut {NoStop}%
\end{thebibliography}%

\clearpage
\pagebreak
\widetext
\begin{center}
\textbf{\large Supplemental Materials: Constraints on Axion-like Particles from a Hard $X$-ray Observation of Betelgeuse}
\end{center}

\setcounter{equation}{0}
\setcounter{figure}{0}
\setcounter{table}{0}
\setcounter{page}{1}
\makeatletter
\renewcommand{\theequation}{S\arabic{equation}}
\renewcommand{\thetable}{S\arabic{table}}
\renewcommand{\thefigure}{S\arabic{figure}}
\renewcommand{\bibnumfmt}[1]{[S#1]}
\renewcommand{\citenumfont}[1]{S#1}


\section{Betelgeuse Stellar Models}
The total ALP number per time and energy can be obtained integrating Eq.~(\ref{eq:axprod}) over the volume of the star, $d\dot N_{a}/dE = \int (d\dot n_{a}/dE)dV$. We find that, with an excellent approximation, the ALP source spectrum has the following form \cite{Andriamonje:2007ew}
\begin{equation}
\frac{d\dot{N}_{a}}{dE} = \frac{10^{42}Cg_{11}^{2}}{\textrm{keV}~\textrm{s}}\left(\frac{E}{E_{0}}\right)^{\beta}e^{-(\beta +1)E/E_{0}} \,\ ,
\end{equation}
where $g_{11}=g_{a\gamma}/10^{-11}\,{\rm GeV}^{-1}$, while $C$ is the normalization, $E_0$ coincides the average energy, and $\beta$ is the spectrum index. The values of $C$, $E_0$ and $\beta$ depend on various structural parameters characterizing the core of the star, such as temperature, density and chemical composition. To this aim, we will make use of stellar models computed using the FuNS code (see \cite{Straniero:2019} for a detailed description of this code and the adopted input physics). Alpha Orionis (Betelgeuse) is a red supergiant whose luminosity, effective temperature and metallicity are, respectively, $\log L/L_\odot=5.10\pm0.22$ (\cite{Bertre:2012bh}), $T_{\rm eff}=3641\pm53\,$K (\cite{Perrin:2004ce}), and $[\mathrm{Fe/H}]=+0.1\pm0.2$ (\cite{Lambert:1984}). These data constrain the initial mass between 18 and 22 $M_\odot$, in good agreement with previous determinations (\cite{Meynet2013, Dolan2016}). We note that the uncertainty of Betelgeuse mass contributes much smaller effect for ALP-photon production than the time to core collapse or $B_{T}$, and is thus ignored here. In our analysis we adopted a model of 20 $M_\odot$ with solar composition.

\par Extant models of stars with mass $\sim 20$ M$_\odot$ may evolve to the red supergiant stage at the onset or at the end of the core-He burning, depending on the assumed efficiency of the semiconvective mixing (see the discussion in section 2.2 in \cite{Straniero:2019}). In the FuNS model, scarce semiconvective mixing is usually assumed, so that the star becomes a red supergiant since the beginning of the core-He burning. This phase lasts for $\sim8\times10^5$\,yrs, during which the central temperature and density remain more or less the same.
After the core He is exhausted, the stellar luminosity increases and attains a maximum value during the C-burning phase. After that, the luminosity remains constant, until the final core collapse. However, during this phase, which lasts a few $10^4$ yrs, the temperature and the density within the core undergo constant and substantial increases. In turn, a significant increase of the ALP-production rate is expected (the ALP production rate is a steep function of the temperature). 

\begin{figure}[h]
    \centering
    \includegraphics[width=0.6\columnwidth]{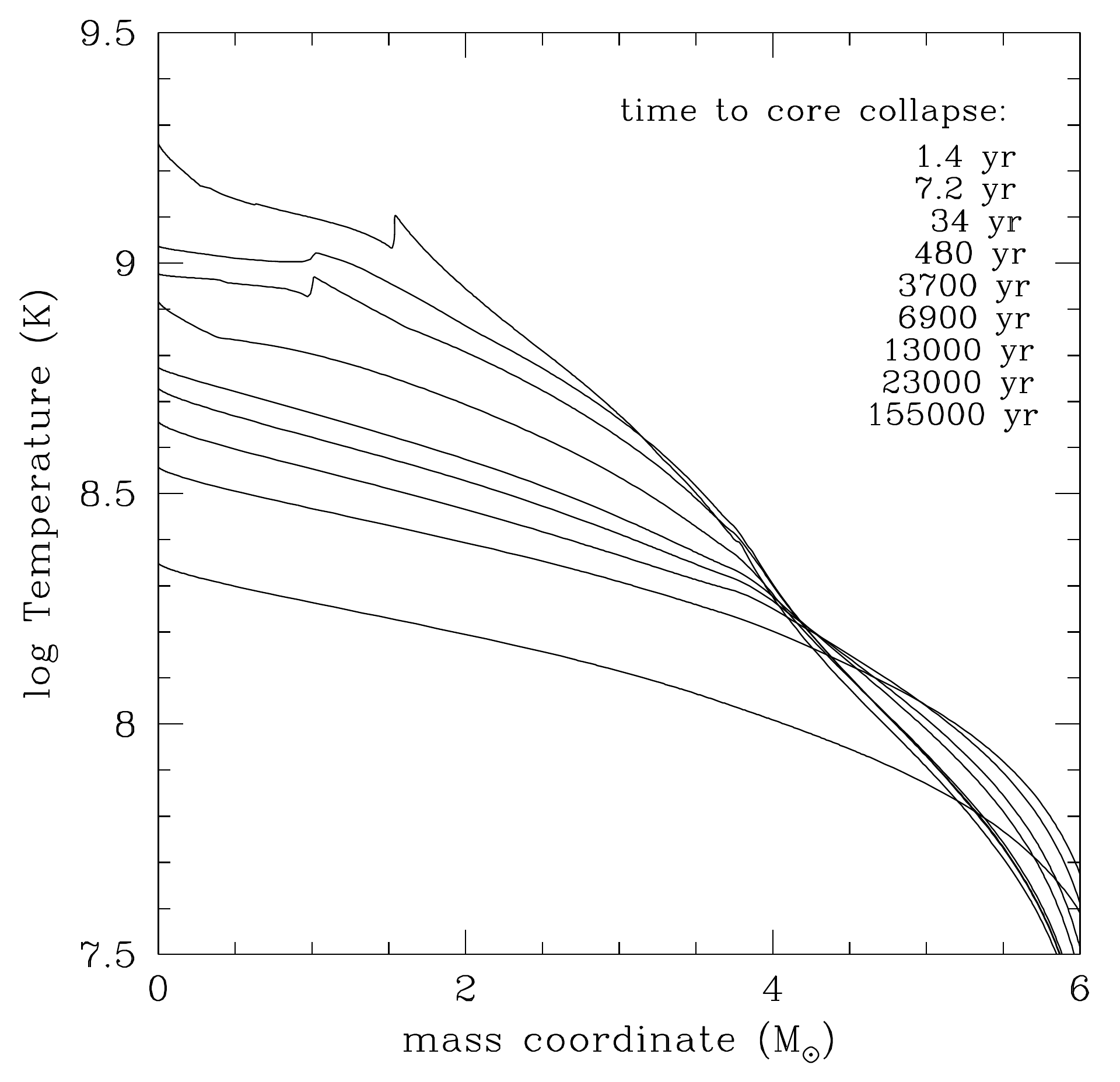}
    \caption{Temperature profiles of some of the 13 models in Tab.~\ref{tab:axion_models}. The plot shows the most internal 6 $M_\odot$ of each model. As the star approaches the final core collapse, the temperature becomes steeper within the core, so that the ALPs production progressively becomes more centrally concentrated. The model labelled 155000 yr to the collapse refers to the He-burning phase, while that 480 yr to the C-burning phase and 1.4 yr to the collapse is taken at the beginning of the O burning. The secondary temperature peaks shown by some models are due to the presence of active C or Ne burning shells.}
    \label{fig:models_tprofile}
\end{figure}

Since the precise evolutionary status cannot be determined from the observed stellar properties, we have considered a set of stellar models taken at different times before the core-collapse. All of these stellar models match the observed $L$ and $T_{\rm eff}$. In this way we may trace the expected evolution of the ALP flux during the red supergiant phase of Betelgeuse. The luminosity, the central temperature and the time to the core collapse of these models are reported in Tab.~\ref{tab:axion_models}.
Model 0 assumes Betelgeuse is still in the core-He burning phase; models 1--4 are before C burning; models 5--9 are during the C burning; model 11 is during the Ne burning; and model 12 is at the beginning of the O burning. Fig.~\ref{fig:models_tprofile} illustrates the temperature profiles of some of the 13 models for the most internal 6 $M_\odot$.

\begin{table}[h!]
\centering
\begin{tabular*}{1.0\columnwidth}{@{\extracolsep{\fill}}cccccccc}
    \hline\hline
    Model & Phase & $t_\mathrm{cc}$ [yr] & $\log_{10}(L_\mathrm{eff}/L_\odot)$ & $\log_{10}(T_\mathrm{eff}/\mathrm{K})$ &  $C$ & $E_{0}$ [keV] &$\beta$ \\
    \hline
    0  & He burning & 155000 & 4.90 & 3.572 & 1.36 & 50 & 1.95 \\
    1  & before C burning & 23000 & 5.06 & 3.552 & 4.0 & 80  & 2.0 \\
    2 & before C burning & 13000 & 5.06 & 3.552 & 5.2 & 99  & 2.0 \\
    3 & before C burning & 10000 & 5.09 & 3.549 & 5.7 & 110 & 2.0 \\
    4 & before C burning & 6900  & 5.12 & 3.546 & 6.5 & 120 & 2.0 \\
    5 & in C burning & 3700  & 5.14 & 3.544 & 7.9 & 130 & 2.0 \\
    6 & in C burning & 730   & 5.16 & 3.542 & 12 & 170 & 2.0 \\
    7 & in C burning & 480   & 5.16 & 3.542 & 13 & 180 & 2.0 \\
    8 & in C burning & 110   & 5.16 & 3.542 & 16 & 210 & 2.0 \\
    9 & in C burning & 34    & 5.16 & 3.542 & 21 & 240 & 2.0 \\
    10 & between C/Ne burning & 7.2   & 5.16 & 3.542 & 28 & 280 & 2.0 \\
    11 & in Ne burning & 3.6   & 5.16 & 3.542 & 26 & 320 & 1.8 \\
    12 & beginning of O burning &1.4   & 5.16 & 3.542 & 27 & 370 & 1.8 \\
    \hline\hline
\end{tabular*}
\caption{Models of ALP production from Betelgeuse. The stage of stellar evolution is parameterized by the time remaining until the core collapse for Betelgeuse, $t_\mathrm{cc}$. See text for the definition of other parameters.}
\label{tab:axion_models}
\end{table}

\begin{figure}[h!]
    \centering
    \subfigure[$m_{a}=1.0\times10^{-11}$ eV (before $\emph{NuSTAR}$)]{\includegraphics[width=0.495\textwidth]{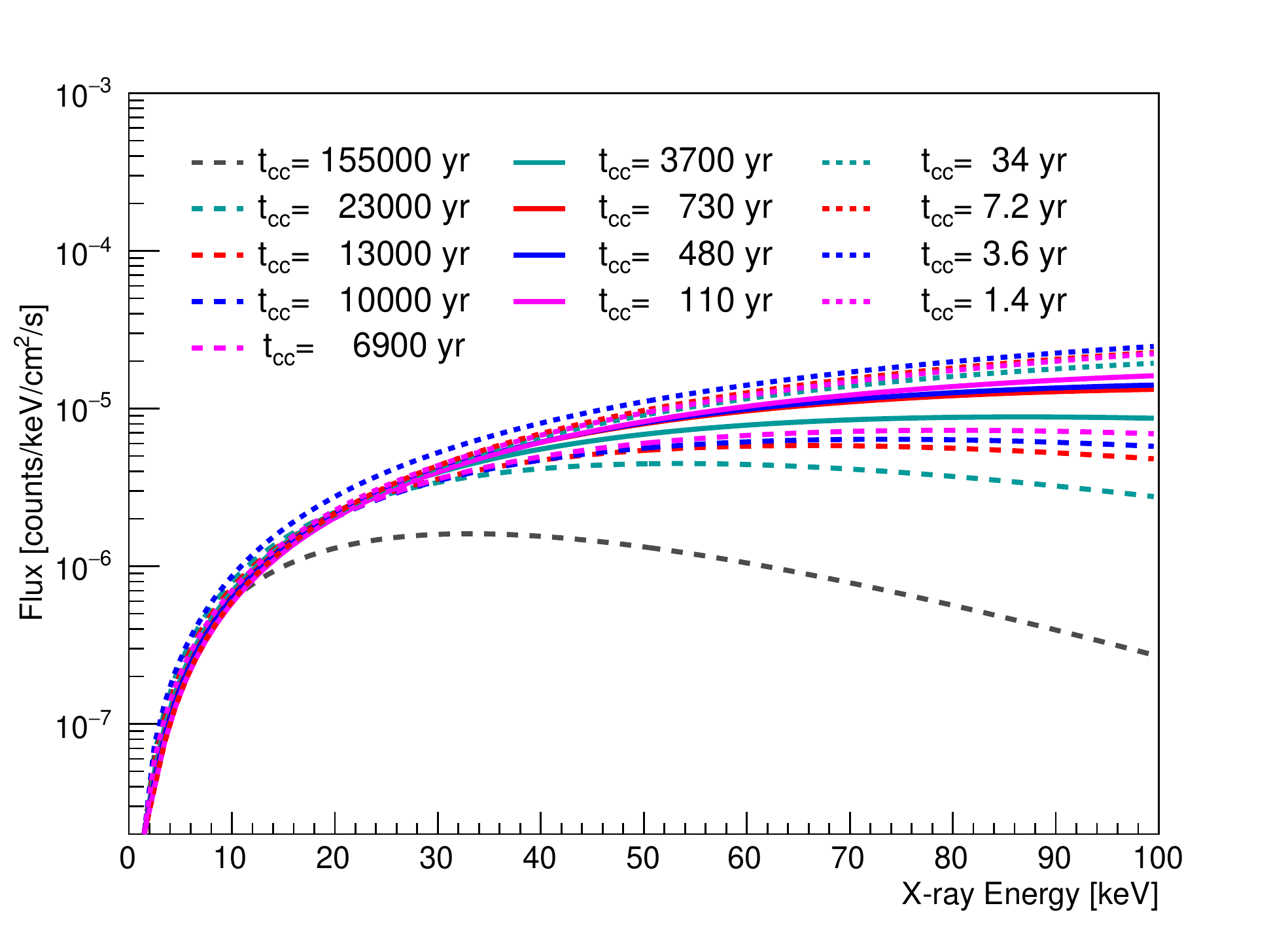}}
    \subfigure[$m_{a}=1.0\times10^{-11}$ eV (after $\emph{NuSTAR}$)]{\includegraphics[width=0.495\textwidth]{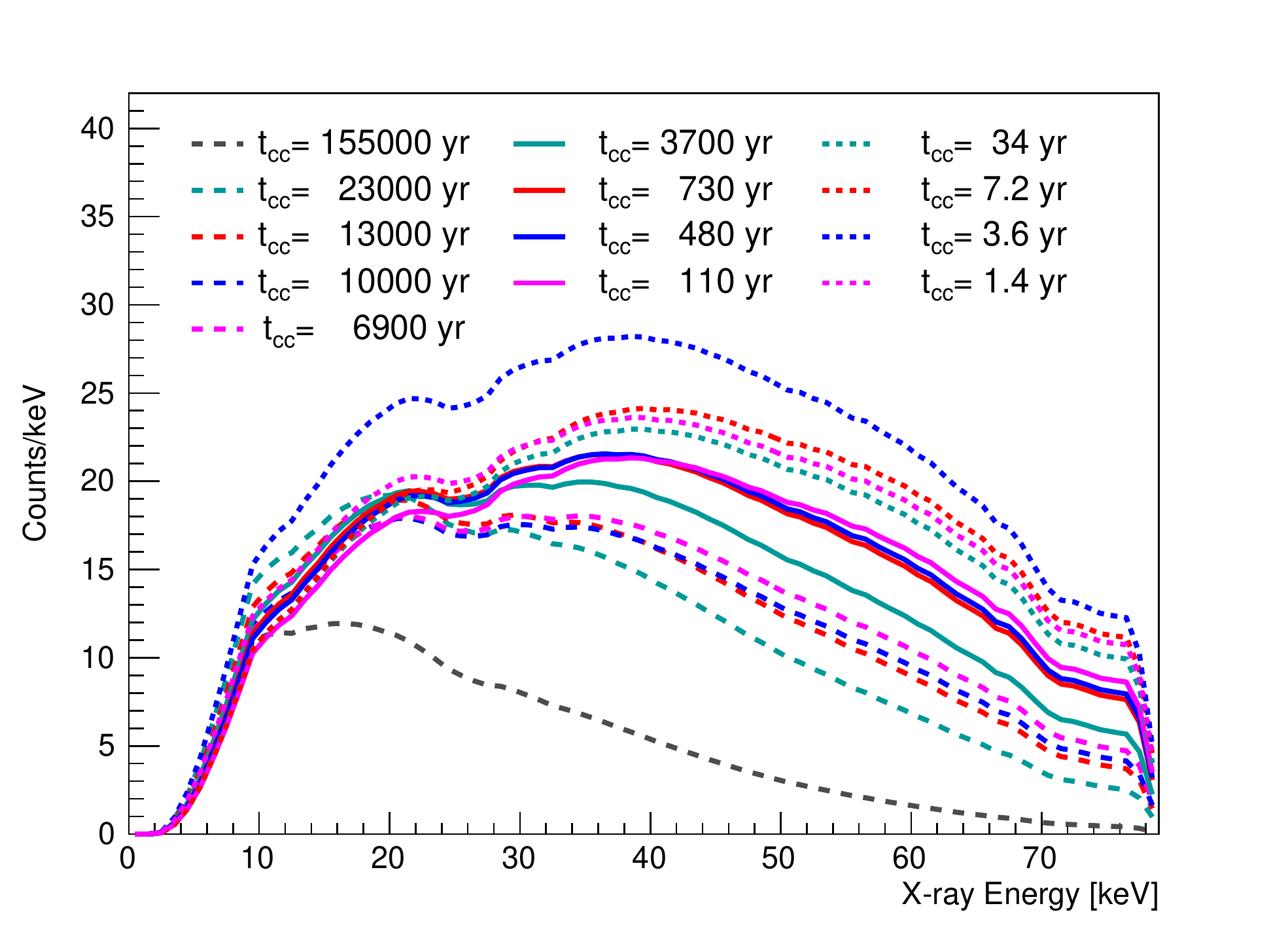}}
    \subfigure[$m_{a}=1.0\times10^{-10}$ eV (before $\emph{NuSTAR}$)]{\includegraphics[width=0.495\textwidth]{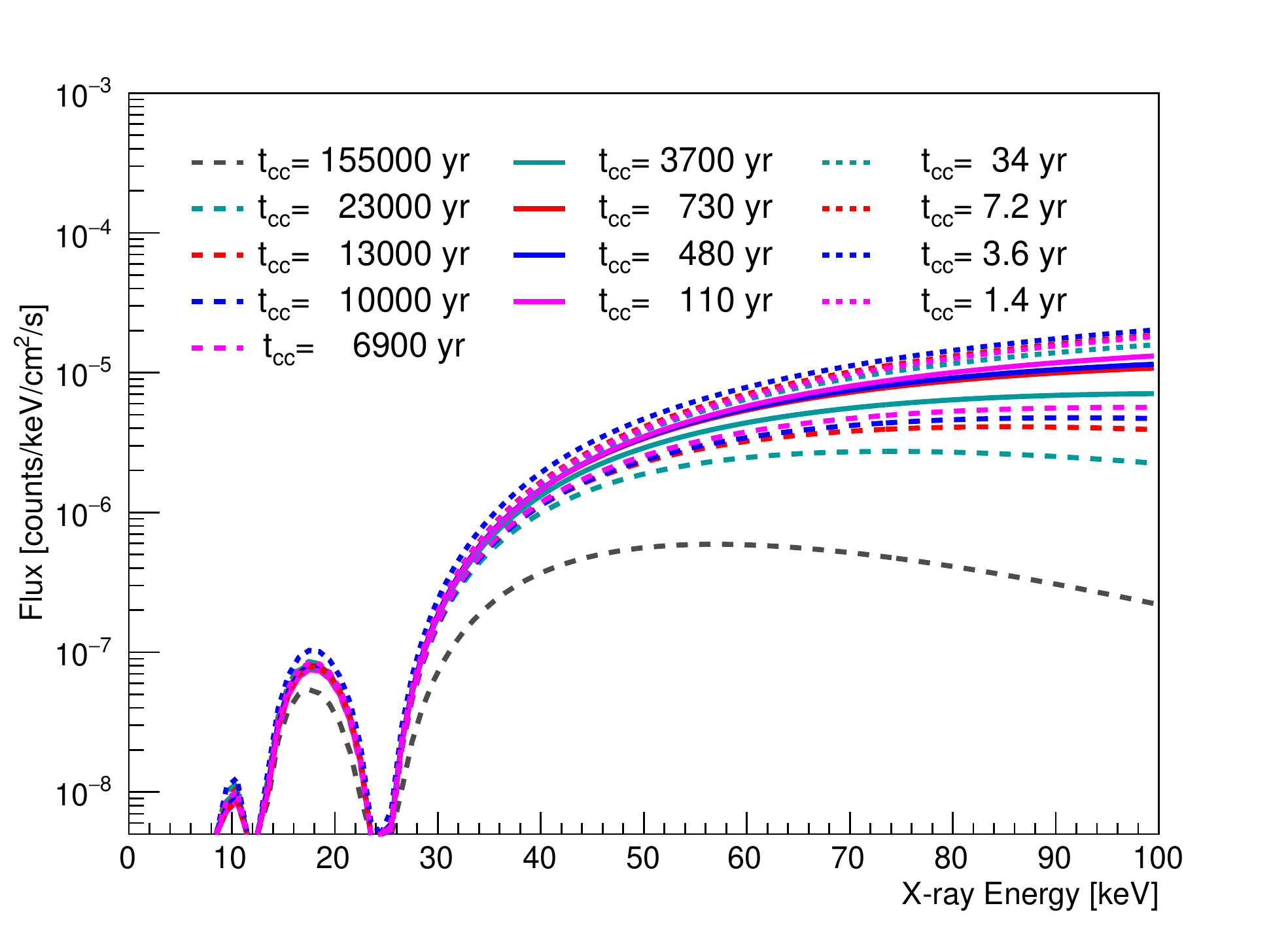}}
    \subfigure[$m_{a}=1.0\times10^{-10}$ eV (after $\emph{NuSTAR}$)]{\includegraphics[width=0.495\textwidth]{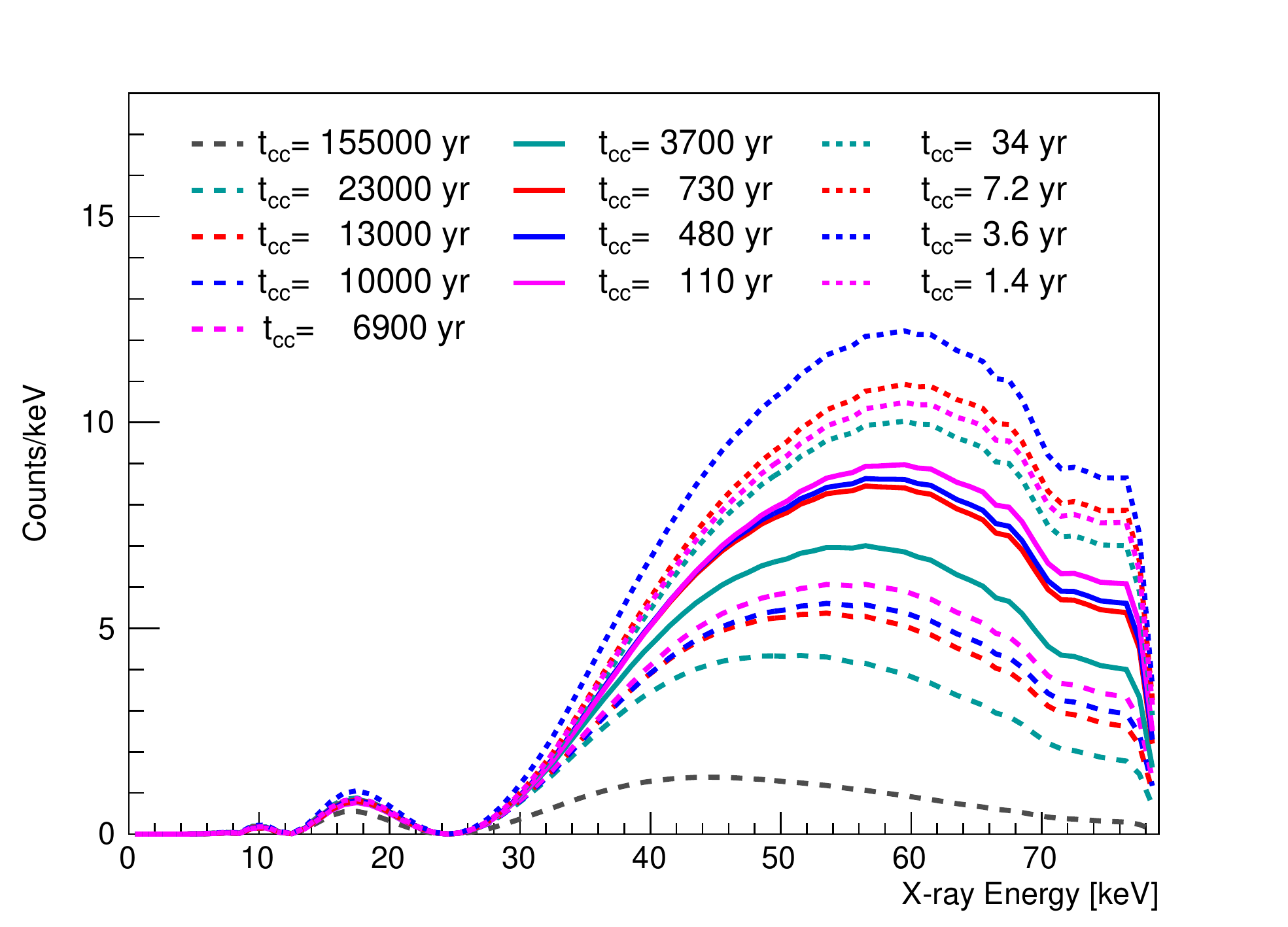}}
    \subfigure[$m_{a}=2.0\times10^{-10}$ eV (before $\emph{NuSTAR}$)]{\includegraphics[width=0.495\textwidth]{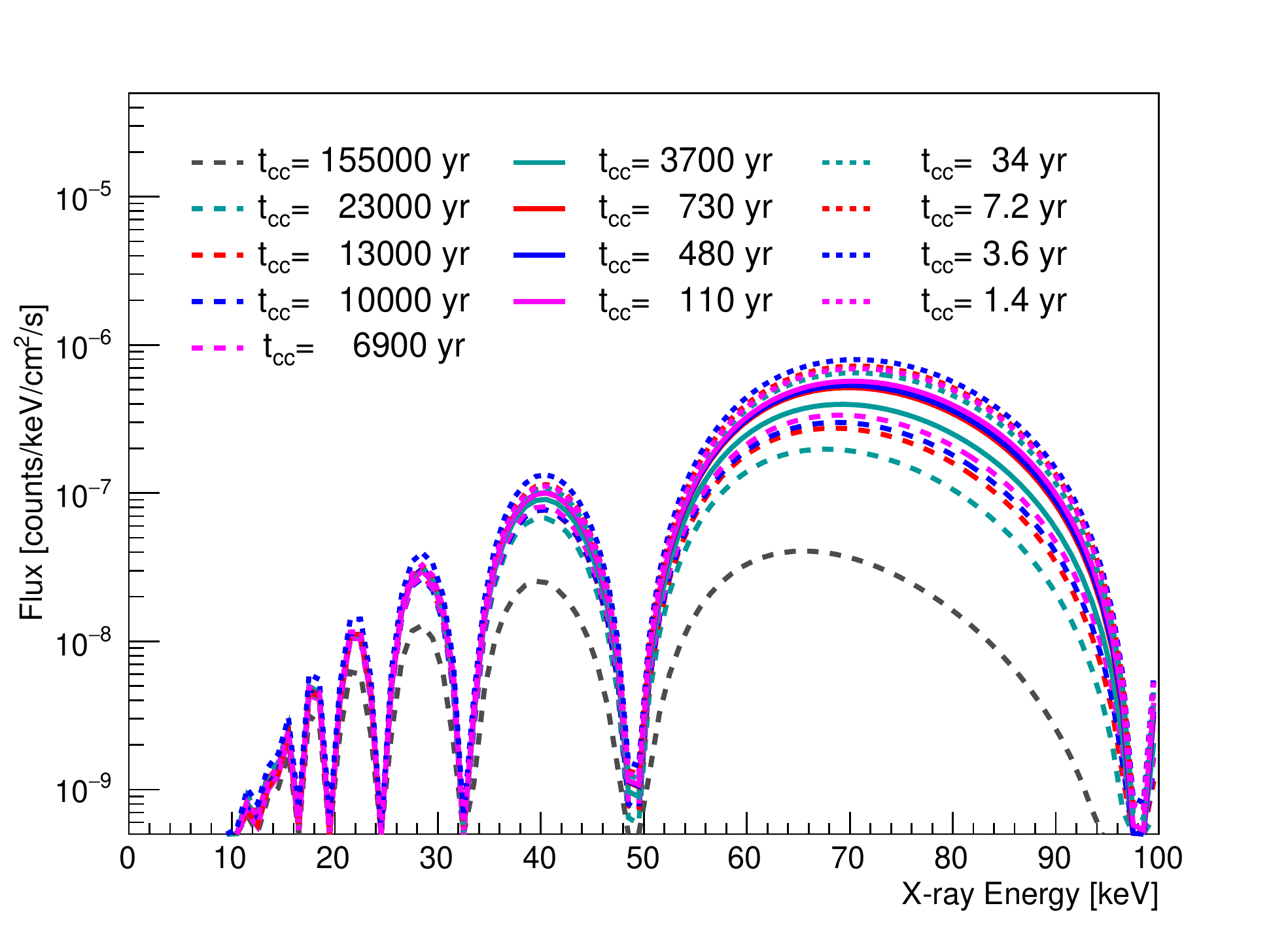}}
    \subfigure[$m_{a}=2.0\times10^{-10}$ eV (after $\emph{NuSTAR}$)]{\includegraphics[width=0.495\textwidth]{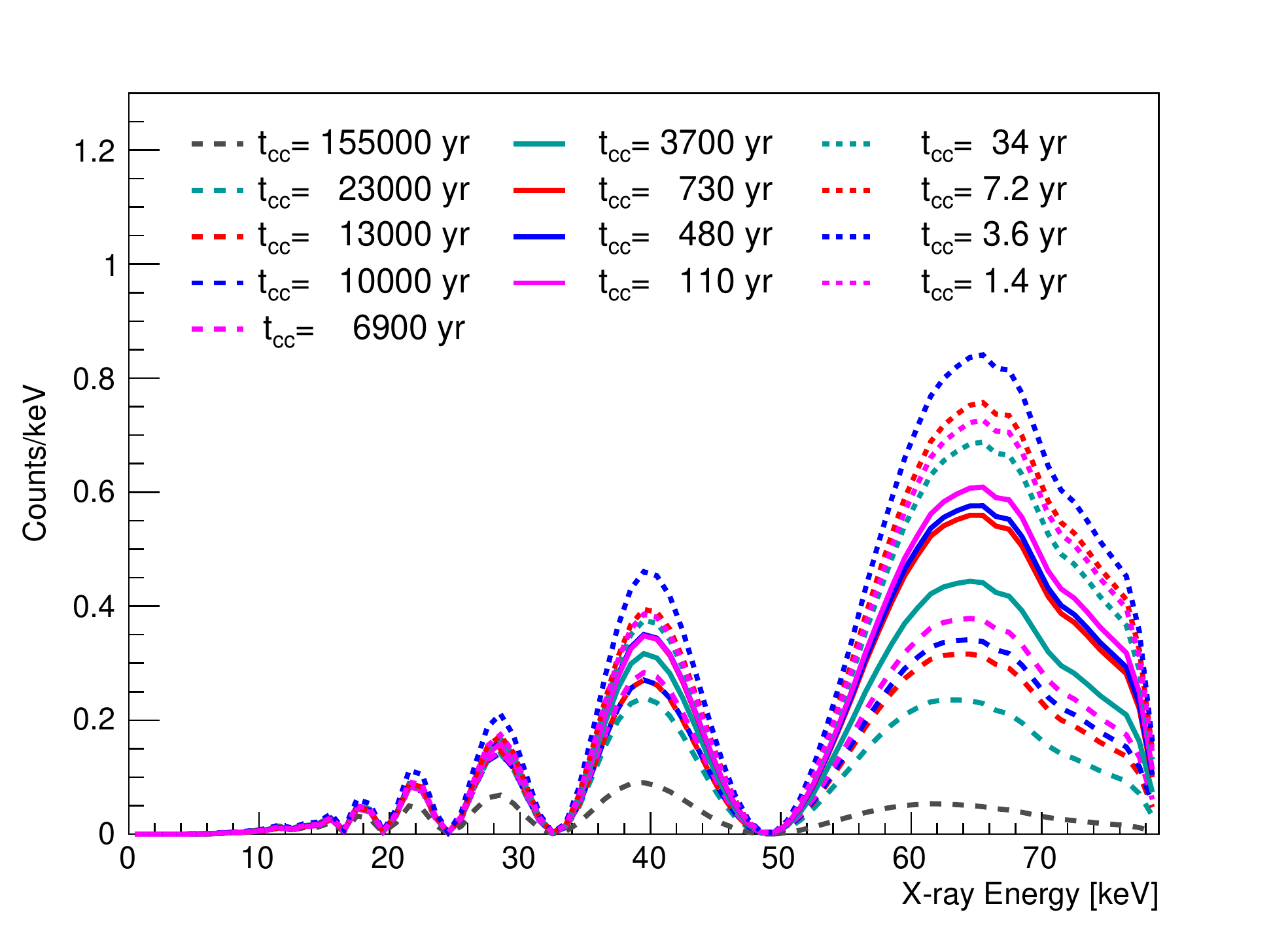}}
    \caption{Predicated $X$-ray spectra \emph{before} and \emph{after} \emph{NuSTAR} instrument response with ALP-photon production models for representative ALP masses with assumption of $B_{T}=1.4~\mu G$ and $g_{a\gamma}=1.5\times10^{-11}~GeV^{-1}$.}
    \label{fig:axion_model_more}
\end{figure}

By folding Eq.~(\ref{eqa:dnr}) from Eq.~(\ref{eq:axprod}), (\ref{eqa:prob}) and (\ref{eqa:qform}), the differential photon flux per unit energy arriving at Earth can be  numerically calculated :
\begin{equation}
\dfrac{d N_\gamma}{d E dS dt} =\dfrac{g_{11}^{4}}{{\rm keV\,cm^{2}\,s}} \left( \dfrac{C}{5.36\times10^{5}} \right) 
\left(\frac{E}{E_0}\right)^\beta e^{-(\beta+1)E/E_0} \left( \dfrac{B_T}{1~\mu G}\right)^2 \left( \dfrac{d}{197~pc} \right)\dfrac{\sin^2 q}{q^2}\,,
\label{eqa:final}
\end{equation}
where 
\begin{equation}
q \simeq \left[ 77\,\left(\dfrac{m_{a}}{10^{-10}\,{\rm eV}}\right)^2-0.14\left( \dfrac{n_e}{0.013~{\rm cm}^{-3}}\right) \right] \nonumber \times \left( \dfrac{d}{197\,{\rm pc}}\right) \left(\dfrac{E}{1\mathrm{\,keV}}\right)^{-1}\, ,
\end{equation}
Using the fitted parameters in Tab.~\ref{tab:axion_models}, Fig.~\ref{fig:axion_model_more} shows the predicated $X$-ray spectra before and after \emph{NuSTAR} instrument response for typical ALP masses in this work.

\newpage
\section{\emph{NuSTAR} Background Modeling}
\par
\begin{figure}[hbt!]
    \centering
    \subfigure[FPMA]{\includegraphics[width=0.45\textwidth]{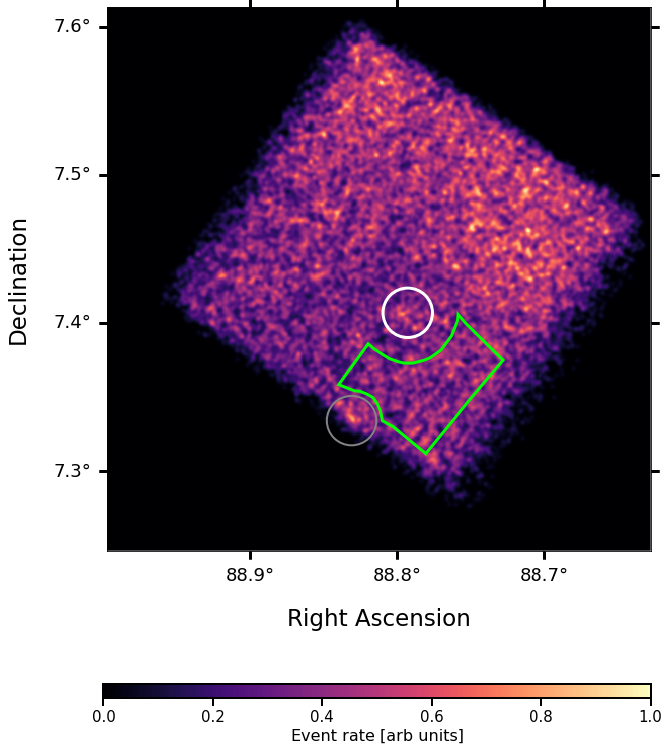}}
    \subfigure[FPMB]{\includegraphics[width=0.45\textwidth]{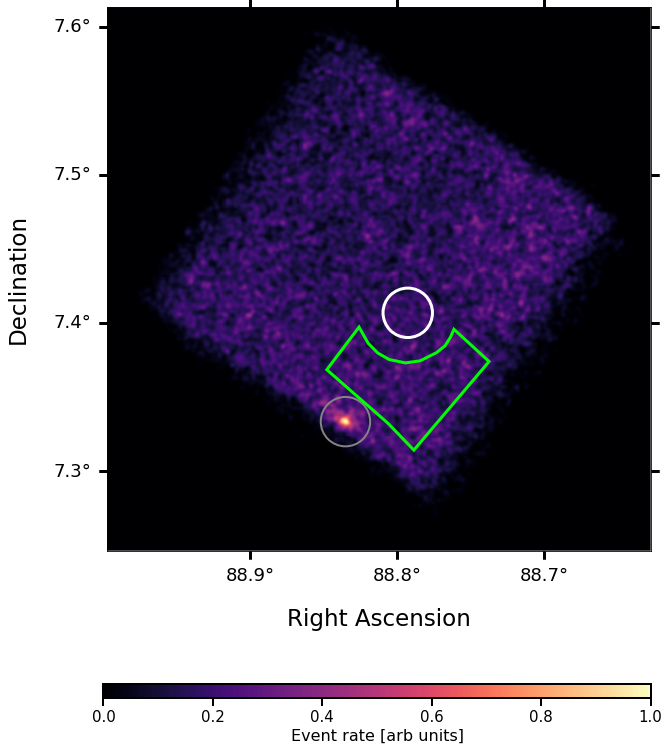}}
    \caption{FPMA (left) and FPMB (right) images of $\emph{NuSTAR}$ observation regions in the energy range 3--79 keV, the event rate in each image is the relative value to the highest one and the image is smoothed with a 2-dimensional Gaussian of width $\sigma = 4.5^{\prime\prime}$ for presentation. The Betelgeuse source region (white circle, 60$^{\prime\prime}$ radius) and background region (green polygon, at least $120^{\prime\prime}$ from Betelgeuse) are shown. The far-away point source is indicated with gray circle (60$^{\prime\prime}$ radius).}
    \label{fig:obs_reg}
\end{figure}

 The \emph{NuSTAR} instrument background spectral model contains several components, which may be broadly categorized as having astrophysical or detector origins. Our ALP search does not require a detailed parametrization of the \emph{NuSTAR} instrument background, so we summarize the most important aspects here and refer the reader to Refs.~\cite{Wik:2014,Perez_2019} for a detailed description. Our primary concern is the uniformity of the instrument background between the source spectral extraction region and the background spectral extraction region. The background components that are known to have significant spatial variation across the detector arrays are the unfocused cosmic $X$-ray background (CXB) and $X$-rays from fluorescence/activation of the instrument structure. The CXB level is observed to be nearly uniform on the sky for the angular and energy acceptance of a single \emph{NuSTAR} observation \cite{Fabian:1992,Revnivtsev:2003wm}, but the shadowing effects of the optics bench and aperture stops produce a radially-varying intensity pattern for the unfocused CXB. The intensity of the detector background is known to vary between the detector chips, though is largely constant within each chip  \cite{Wik:2014}. Both the unfocused CXB gradient and the detector emission motivate the choice of a background region as close as possible to---and ideally on the same detector chip as---the source region, as shown in Fig.~\ref{fig:obs_reg}.

\par
\begin{figure}[h]
    \centering
    \subfigure[FPMA]{\includegraphics[width=0.49\textwidth]{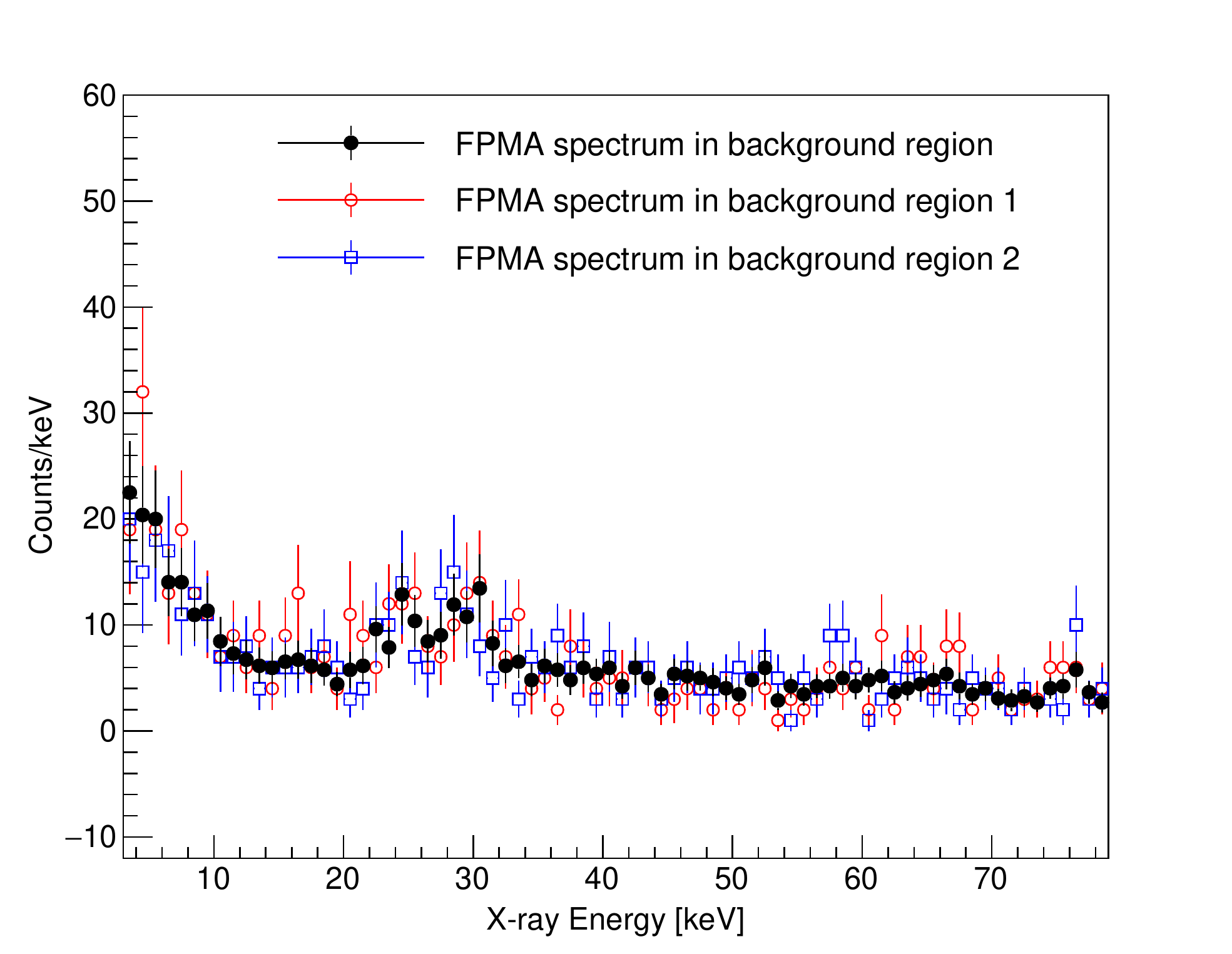}}
    \subfigure[FPMB]{\includegraphics[width=0.49\textwidth]{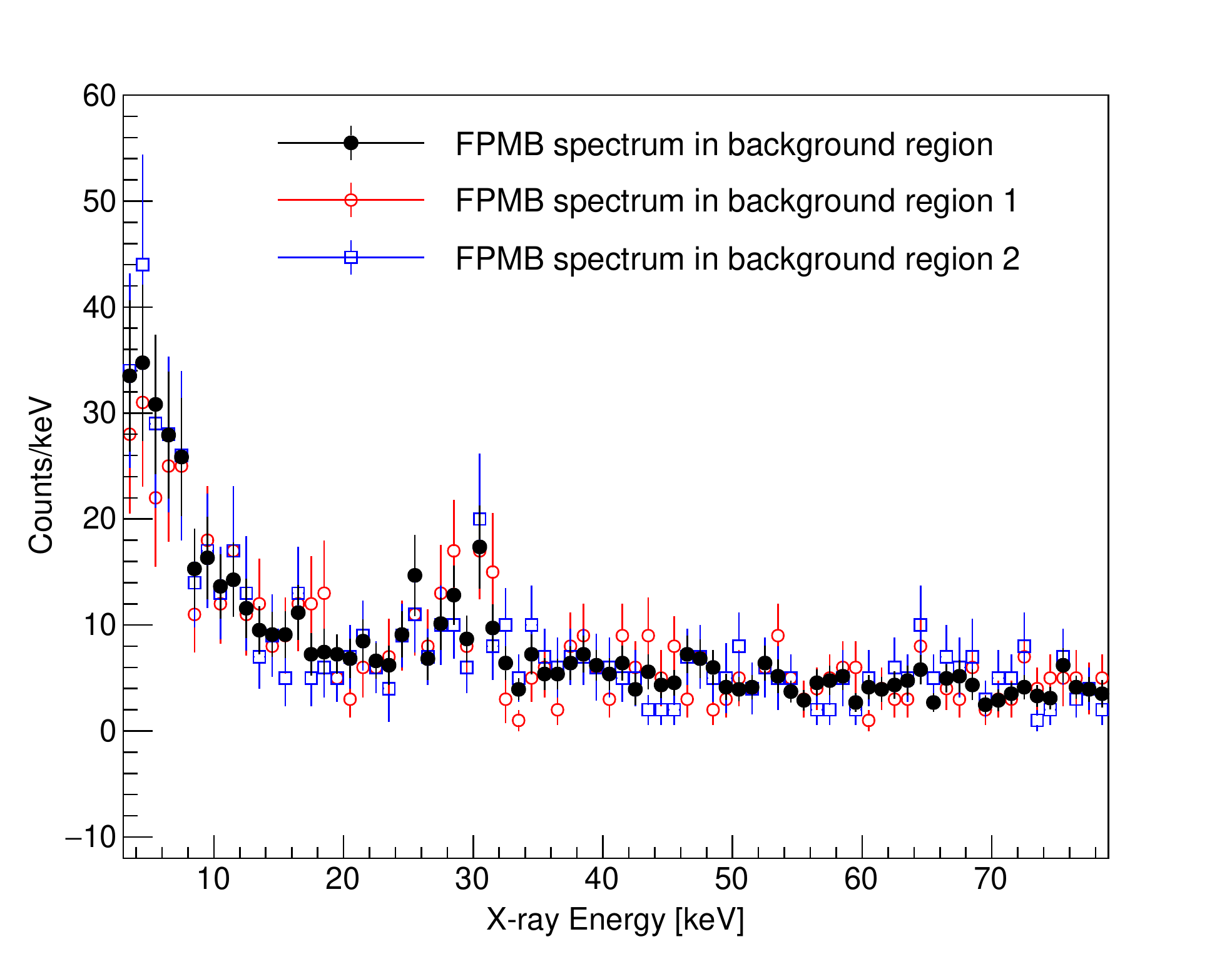}}
    \caption{$X$-ray spectra in 3--79 keV energy range from FPMA and FPMB in different background regions after the normalization of region size: the black is the spectrum from the polygon region shown in Fig.~\ref{fig:obs_reg}; the red and blue are for the other region choices (see text for details). The data is binned with the width of 1 keV for presentation. The error bars are are calculated by Sumw2 with ROOT. }
    \label{fig:bkgspec_regs}
\end{figure}

 We further confirm the spatial uniformity of background by comparing the $X$-ray spectra with different choices of background region on the same detector chip. In addition to the polygon region shown in Fig.~\ref{fig:obs_reg}, we also choose the other two circle regions (60$^{\prime\prime}$ radius) which are 120$^{\prime\prime}$ away from the Betelgeuse center. Using the data process described above, we extract the $X$-ray spectrum in each background region. Fig.~\ref{fig:bkgspec_regs} compares the spectra from three background regions after the normalization of region size for FPMA and FPMB. This confirms that our analysis is robust to the exact choice of background spectral extraction region, and that our background region accurately models the instrumental and astrophysical background underlying our source region.

\section{\emph{Chandra} Observation of Betelgeuse}
\par
\begin{figure}[t]
\centering
\includegraphics[width=0.5\columnwidth]{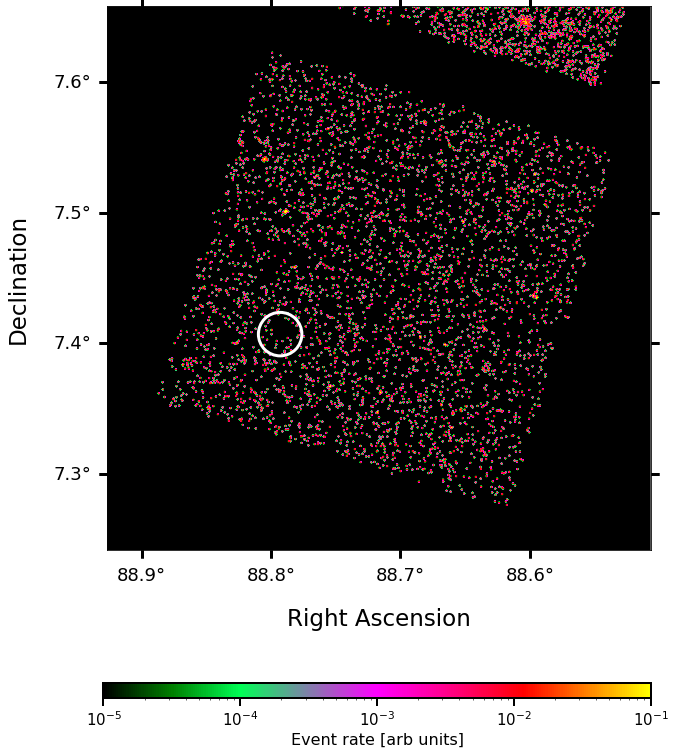}
\caption{Observation image for $0.3-8$ keV soft $X$-ray events from \emph{Chandra}, the event rate is the relative value to the highest one and the image is smoothed with a 2-dimensional Gaussian of width $\sigma = 4.5^{\prime\prime}$ for presentation.The Betelgeuse source region is indicated with the white circle.}
\label{fig:img_chandra}
\end{figure}

 As a cross-check of our results, we use a 5 ks \emph{Chandra} observation of Betelgeuse. \emph{Chandra} has lower, and better understood, instrument background than \emph{NuSTAR}, but its low-energy range ($<$10 keV) makes it less powerful for constraining the expected ALP-induced spectrum. We thus use the Chandra data to verify that our derived 95\% C.L. on $g_{a\gamma}$ is consistent with this low-background, low-energy dataset. We analyzed an archival $\textit{Chandra}$ observation, ObsID 3365, which used the ACIS-I configuration in FAINT mode and was taken on 16 December 2001. We reprocessed the data using the standard $\tt{chandra\_repro}$ tool from the CIAO v.4.12 software, with updated calibration files (CALDB v4.9.2.0, \cite{Fruscione06}). The resultant cleaned data had a 4.899~ks exposure. Fig.~\ref{fig:img_chandra} shows the observation image for $0.3-8$ keV soft $X$-ray events from \emph{Chandra}. The short exposure time and low count rates made analyzing background-subtracted spectra difficult. Rather, we used the new CIAO $\tt{aprates}$ tool, which computes values and limits for various parameters such as point source count rate and photon flux (\url{https://cxc.harvard.edu/ciao/threads/aprates/index.html}).
\par Upper limits for the $0.3-2.5$~keV energy range emission from Betelgeuse for ObsID 3365 were first reported by Posson-Brown et al. (2007) (see Table 4 there). We used the same source ($r=4^{\prime\prime}$) and background ($r_{min}=10^{\prime\prime}$, $r_{max}=10^{\prime\prime}$) regions centered on the star's equatorial position, and with $\tt{aprates}$ calculated a $3\sigma$ upper limit that matched the reported upper limit count rate. We then extended our analysis to the $0.3-8$~keV energy range. The resultant upper limits are 4.6 counts, or $9.42\times10^{-4}$ counts~s$^{-1}$. By counting the total events in the energy range $0.3-8$ keV, the constrains on $g_{a\gamma}$ is set as $2.2 \times10^{-11}~GeV^{-1}$ for the most optimistic stellar model and $B_{T} = 3~\mu G$, and $6.5\times10^{-11}~GeV^{-1}$ for the most conservative stellar model and $B_{T}=0.4~\mu G$ for $m_{a}<3.5\times10^{-11}~eV$. These values are well within our excluded regions, and thus we confirm that our results are consistent with this low-energy dataset.

\section{Constraints on $g_{a\gamma} \times \sqrt{B_{T}}$}
\begin{figure}[htb!]
\centering
\includegraphics[width=0.75\columnwidth]{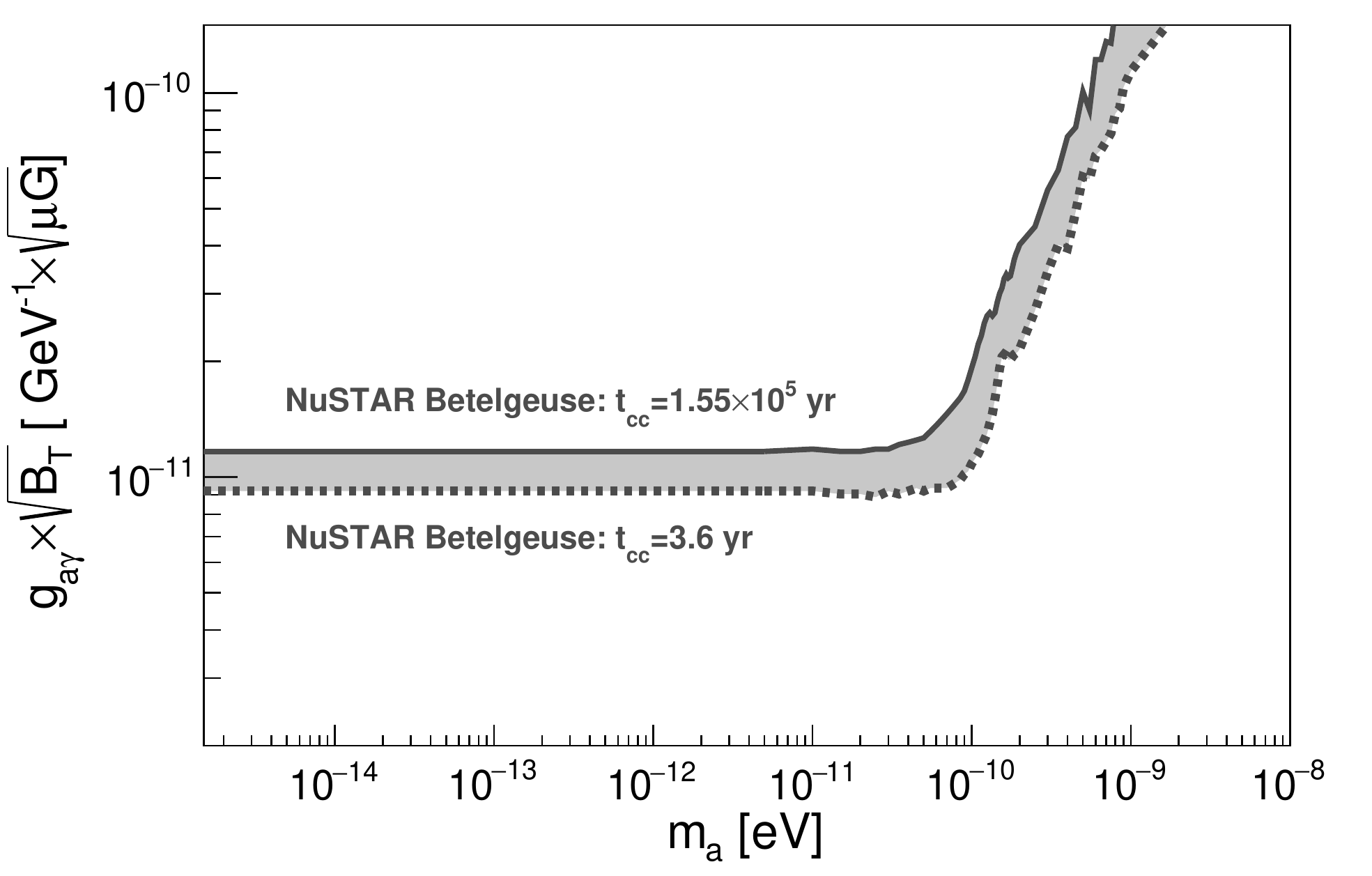}
\caption{Upper limits on $g_{a\gamma} \times \sqrt{B_{T}}$: the solid line is for the most conservative stellar model, dashed line for the most optimistic stellar model and the band for the other models between in.}
\label{fig:results_garbt}
\end{figure}

As shown in Eq.~\ref{eqa:final}, the ALP-photon flux scales as $g_{a\gamma}^{4} \cdot B_\mathrm{T}^{2}$. To separately discuss the uncertainty from Betelgeuse stellar model ($t_{cc}$) and magnetic field ($B_{T}$), we also present our results as the constraints on the production of $g_{a\gamma}$ and $\sqrt{B_{T}}$ in Fig.~\ref{fig:results_garbt}. We set a constraint of $1.2\times 10^{-11} GeV^{-1} \sqrt{\mu G}$ for $m_{a}<3.5\times10^{-11}$ eV for $t_{cc}=1.55\times10^{5}$ yr, and $0.9 \times 10^{-11} GeV^{-1} \sqrt{\mu G}$ for $m_{a}<5.5\times10^{-11}$ eV for $t_{cc}=3.6$ yr.

\section{Evolution of $g_{a\gamma}$ for more ALP masses}
\begin{figure}[h]
    \centering
    \subfigure[$m_{a}=1.0\times10^{-11}$ eV]{\includegraphics[width=0.435\textwidth]{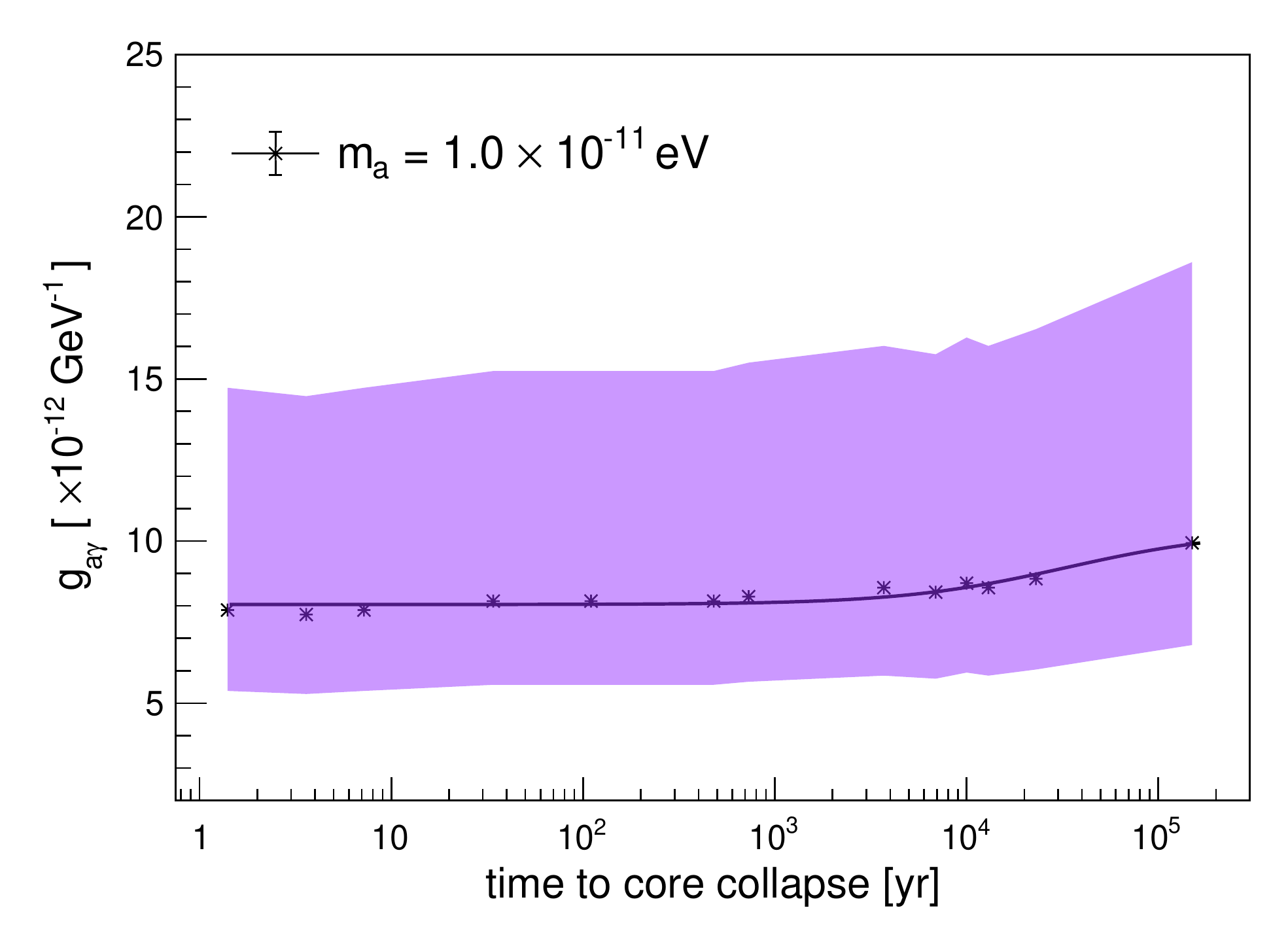}}
    \subfigure[$m_{a}=5.0\times10^{-11}$ eV]{\includegraphics[width=0.435\textwidth]{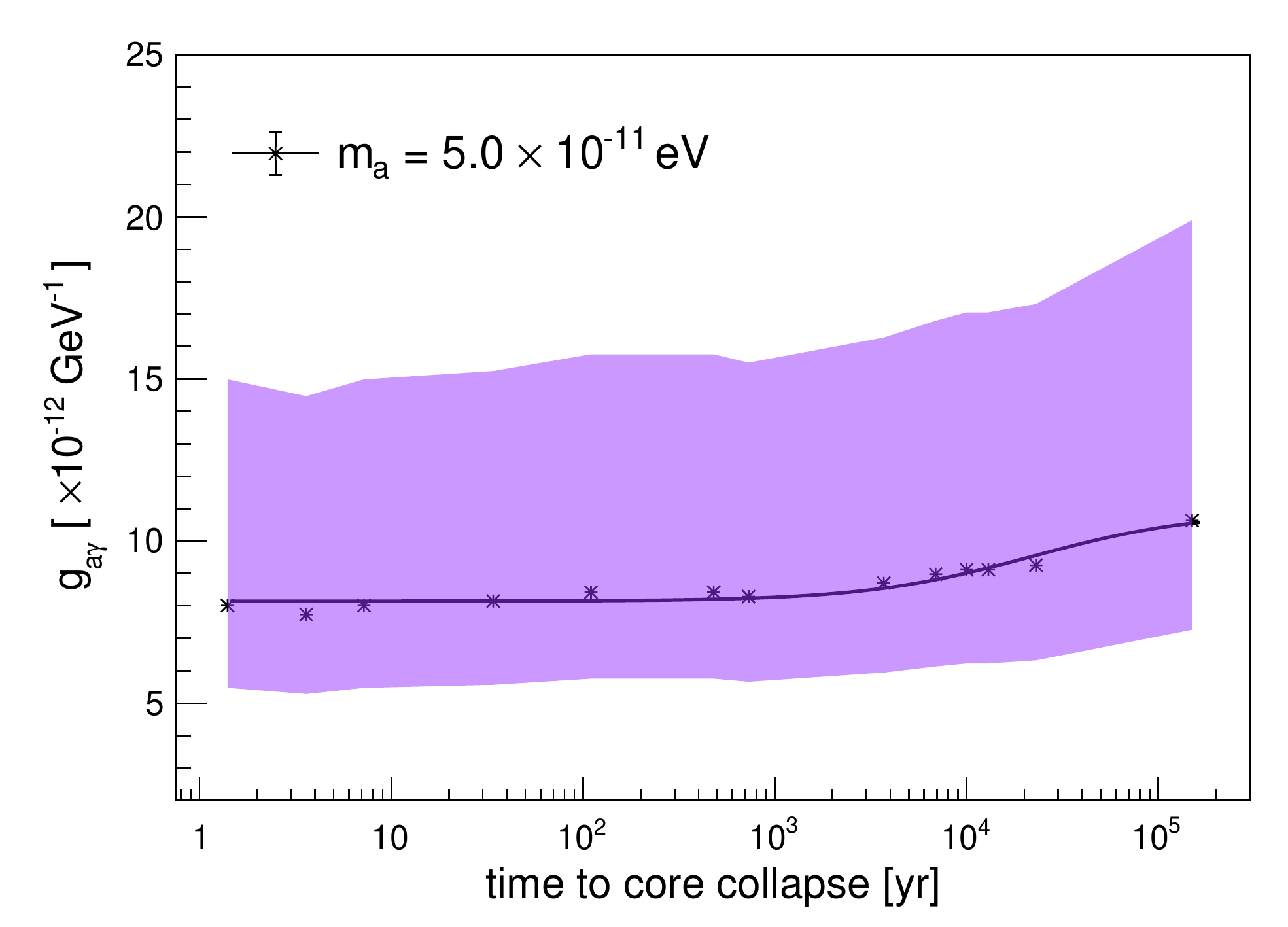}}
    \subfigure[$m_{a}=1.0\times10^{-10}$ eV]{\includegraphics[width=0.435\textwidth]{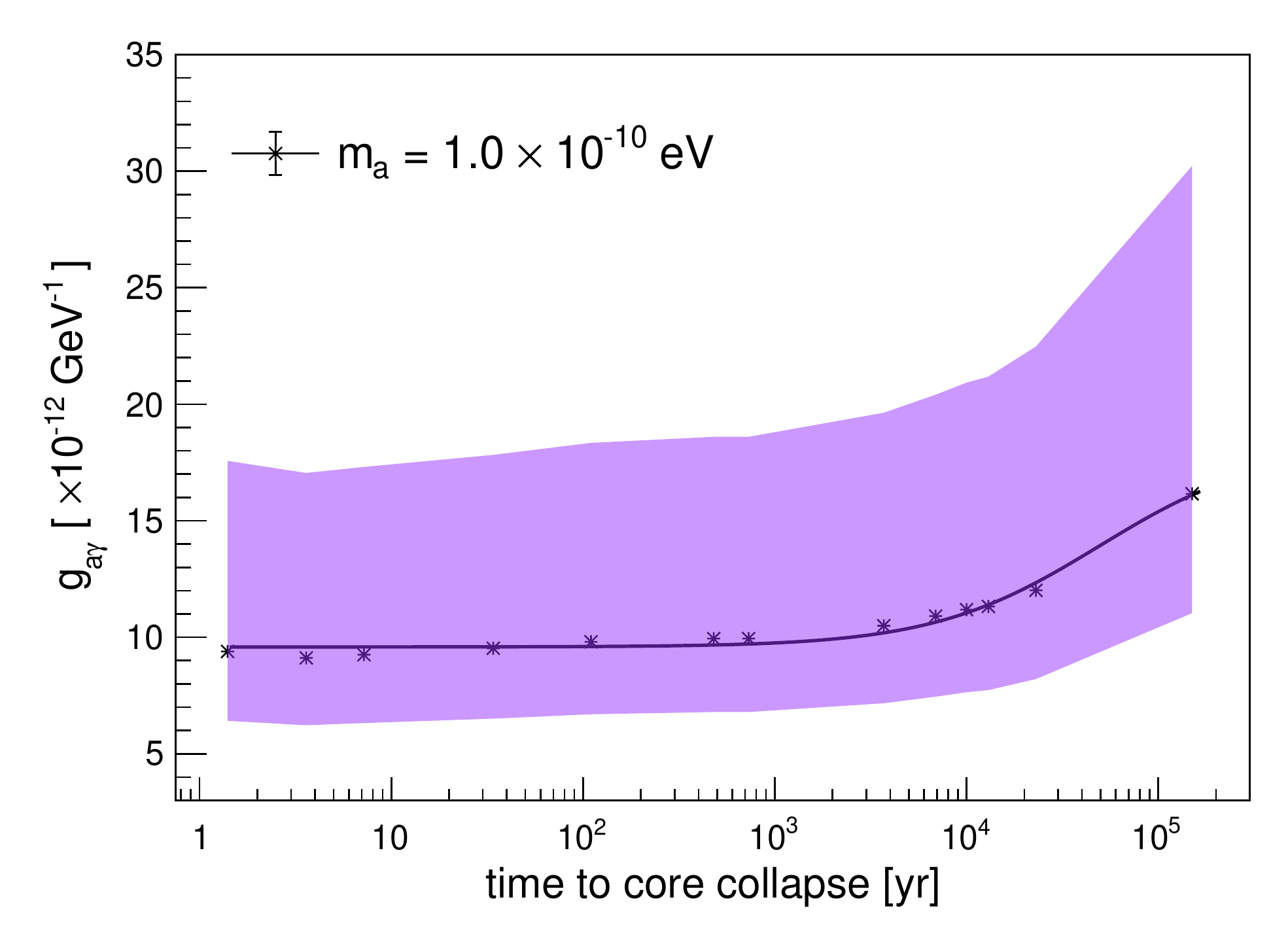}}
    \subfigure[$m_{a}=1.5\times10^{-10}$ eV]{\includegraphics[width=0.435\textwidth]{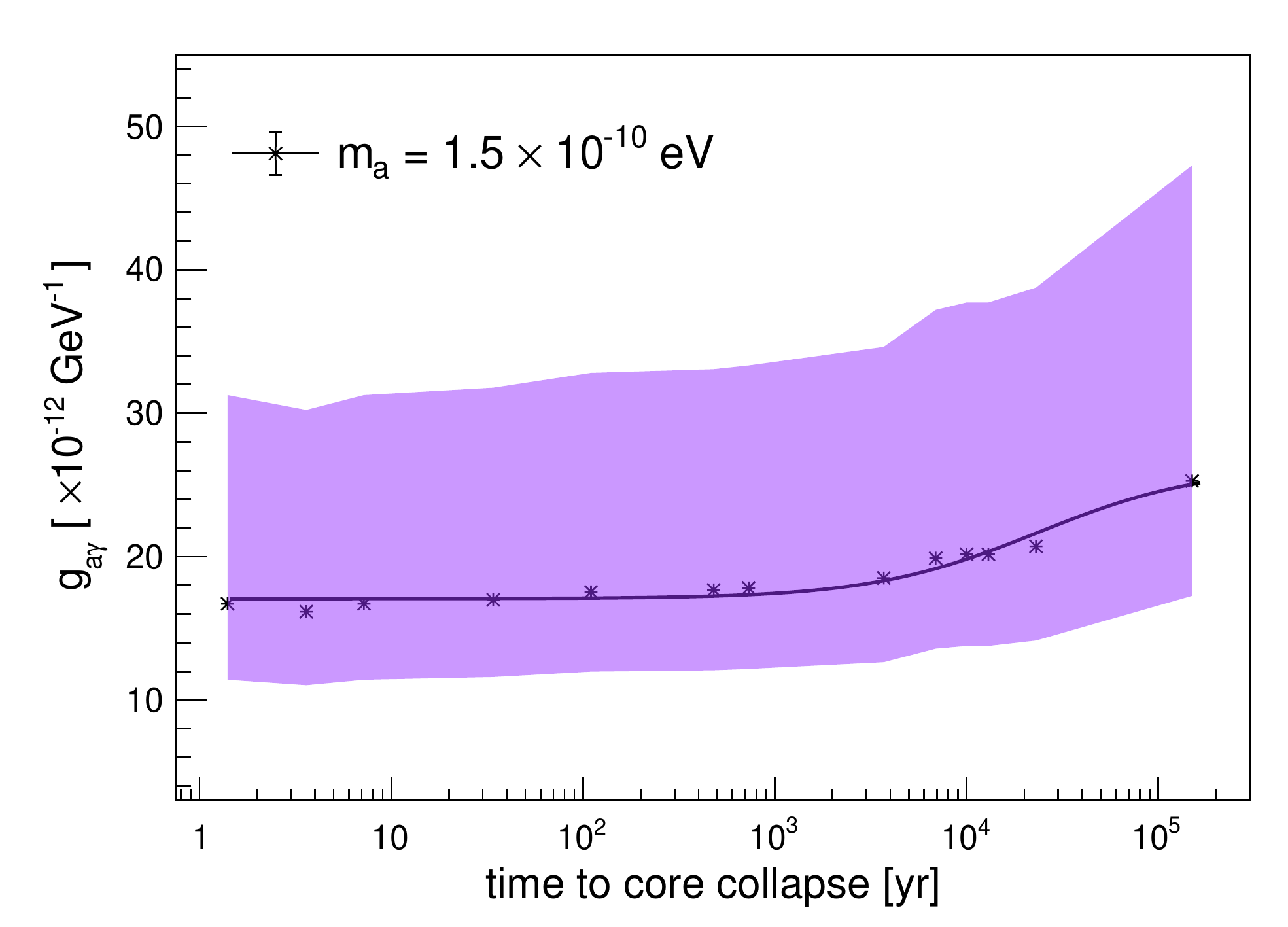}}
    \subfigure[$m_{a}=2.0\times10^{-10}$ eV]{\includegraphics[width=0.435\textwidth]{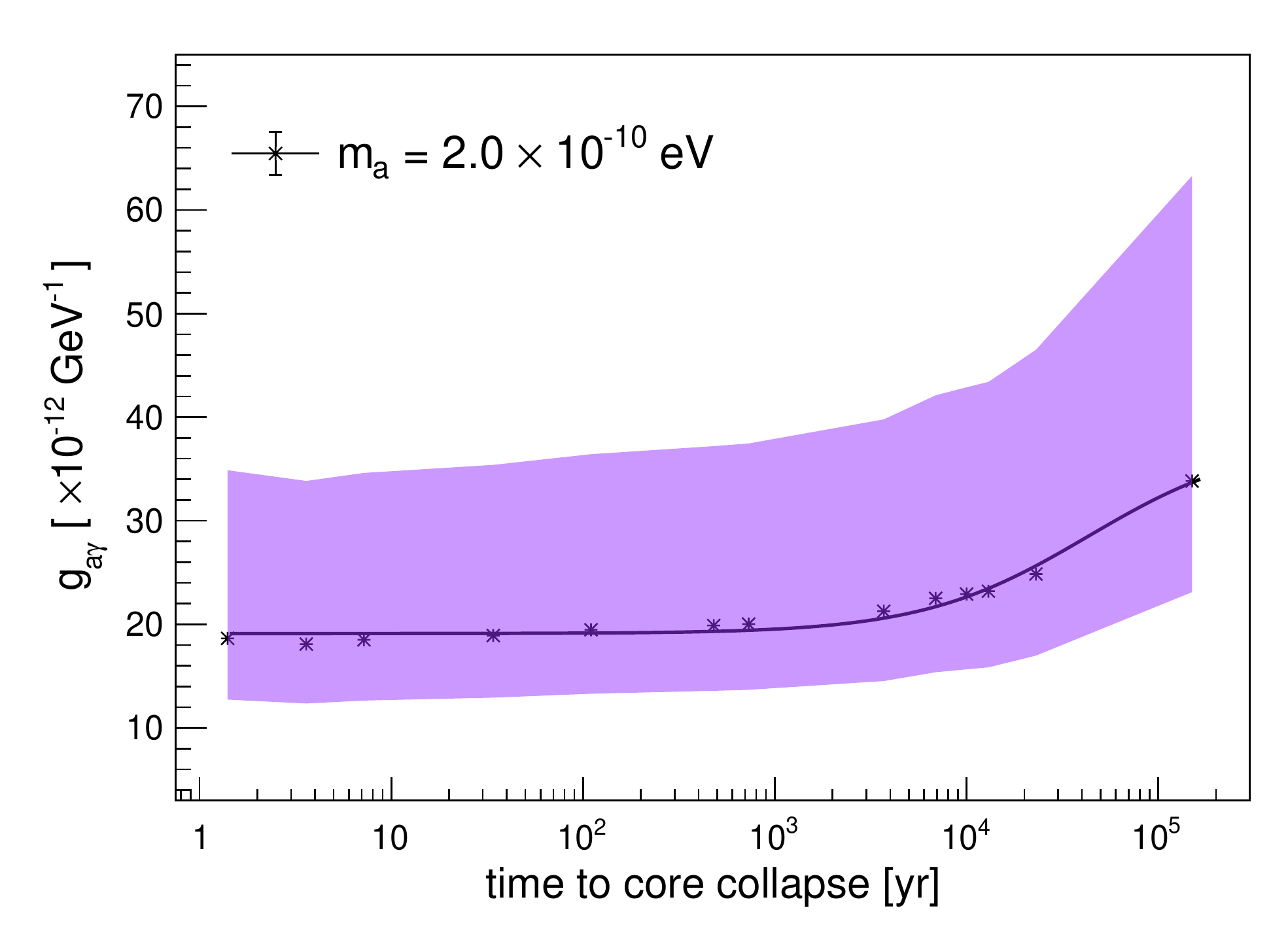}}
    \subfigure[$m_{a}=5.0\times10^{-10}$ eV]{\includegraphics[width=0.435\textwidth]{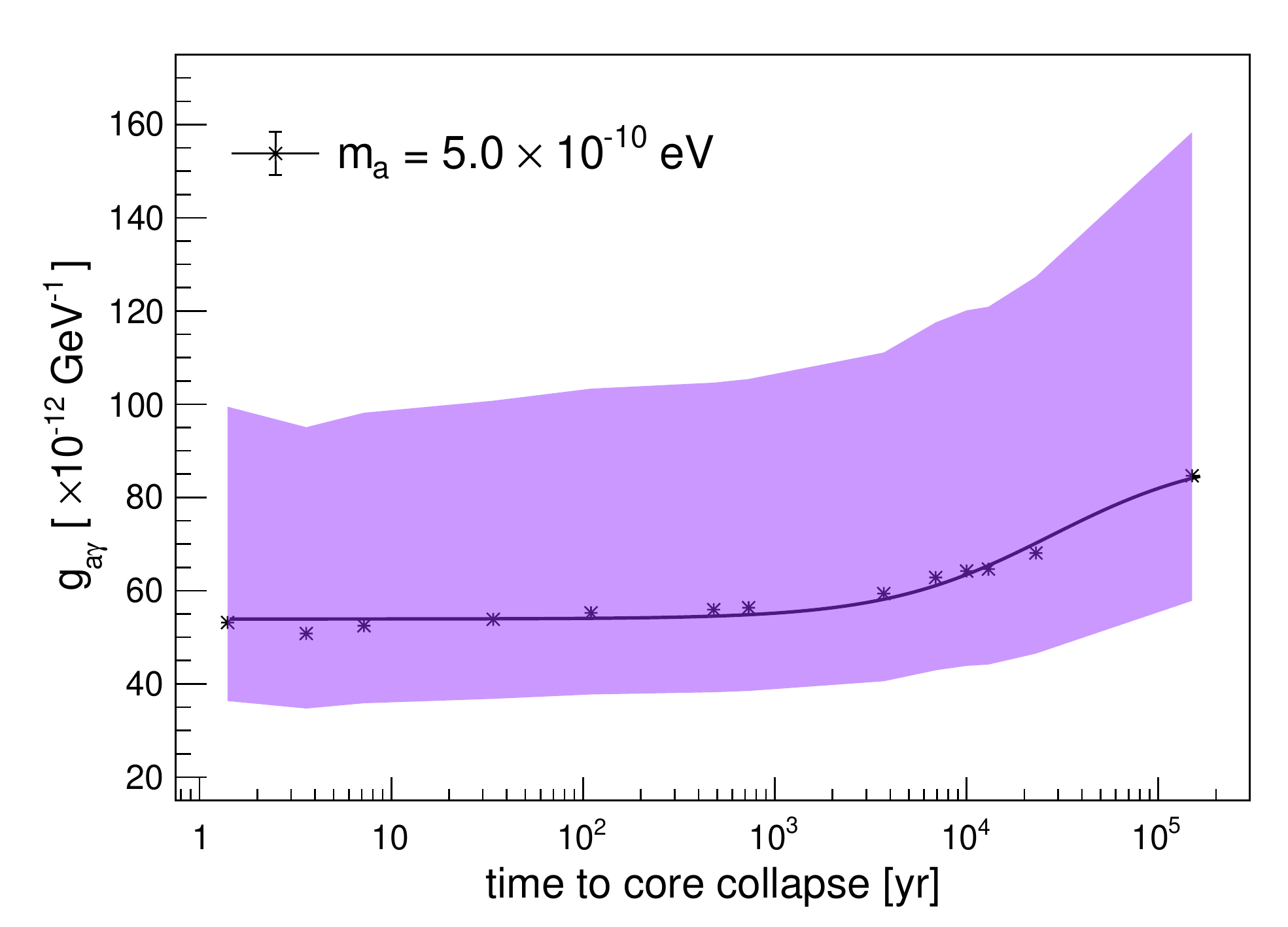}}
    \caption{Evolution on the derived 95\% C.L. upper limit of $g_{a\gamma}$ for different ALP masses with remaining time until core collapse for Betelgeuse. The points are the results with the assumption of $B_{T}=1.4~\mu G$ and the solid black line shows the fitting. The width of the violet band indicates the uncertainty due to choice of $B_{T}$.}
\label{fig:gar_evolution_more}
\end{figure}

Fig.~\ref{fig:gar_evolution_more} illustrates the $g_{a\gamma}$ evolution as the remaining time for core collapse of Betelgeuse for more ALP masses.

\end{document}